\documentclass[11pt]{article}
\pdfoutput=1
\usepackage{appendix}

\usepackage{epsfig}
\usepackage{graphicx}
\usepackage{verbatim}

\usepackage{caption}

\usepackage{latexsym,times}
\usepackage{amsmath}
\usepackage{amsxtra}
\usepackage{amscd}
\usepackage{amsthm}
\usepackage{amsfonts}
\usepackage{amssymb}
\usepackage{dsfont}
\usepackage{xcolor}
\usepackage{cancel}
\usepackage{framed}
\usepackage[normalem]{ulem}

\usepackage{enumitem}
\usepackage{float}

\usepackage{hyperref}
\usepackage[square,sort&compress,comma,numbers]{natbib}

\makeatletter

\@addtoreset{equation}{section} \makeatother

\newcommand\nn{\nonumber}

\newcommand\So{S_{\scriptscriptstyle{{(0)}}}}
\newcommand\To{\EM_{\scriptscriptstyle{{(0)}}}}

\newcommand{\half}{{{\textstyle\frac{1}{2}}}}
\newcommand{\quarter}{{{\textstyle\frac{1}{4}}}}
\newcommand{\be}{\begin{equation}}
\newcommand{\ee}{\end{equation} }
\newcommand{\beqa}{\begin{eqnarray} }
\newcommand{\eeqa}{\end{eqnarray} }
\newcommand{\ba}{\begin{array}}
\newcommand{\ea}{\end{array}}
\newcommand{\bpm}{\begin{pmatrix}}
\newcommand{\epm}{\end{pmatrix}}

\newcommand{\Spin}{\mathbf{Spin}}

\newcommand{\rmd}{{\rm d}}

\newcommand{\ODD}{\mathbf{O}(D,D)}

\newcommand\EM{T}
\newcommand\EK{K}

\newcommand\diag{\mbox{diag}}

\newcommand{\YM}{\rm{YM}}
\newcommand{\DFT}{\rm{DFT}}

\newcommand\rd{{\rm d}}

\newcommand\cA{{\cal A}}

\newcommand\cD{{\cal D}}
\newcommand\cE{{\cal E}}

\newcommand\cH{{\cal H}}

\newcommand\cJ{{\cal J}}

\newcommand\cL{{\cal L}}

\newcommand\fcL{{{\widetilde{\cal L}}}}

\newcommand\hrho{\hat{\rho}}

\newcommand\dis{\displaystyle}

\def\tpartial{\tilde{\partial}}

\def\bra{\bar{a}}

\def\bre{\bar{e}}

\def\breta{\bar{\eta}}

\def\brpsi{\bar{\psi}}

\def\brp{{\bar{p}}}
\def\brq{{\bar{q}}}

\def\brV{{\bar{V}}}

\def\brP{{\bar{P}}}

\newcommand{\na}{{\nabla}}
\newcommand{\trd}{{\bigtriangledown}}

\newcommand\p\partial

\def\HE{H_{\textrm{E}}}

\begin{document}
\begin{titlepage}
\title{\vspace{-3.7cm}\bf $\mathbf{O}(D,D)$ completion of the Friedmann equations}
\author{\sc Stephen Angus,${}^{1}$\,   \,Kyoungho Cho,${}^{2}$\,  \,Guilherme Franzmann,${}^{3,4}$\\ \sc \,Shinji Mukohyama,${}^{4,5}$   \,\,and\,\, \,Jeong-Hyuck Park\,${}^{2,4}$\footnote{On sabbatical leave from 2.}}

\date{}
\maketitle 
\vspace{-1.0cm}
\begin{center}
\mbox{\!\!\!\!\!\!\!\!\!\!\!\!\!\!\!\!\!\!\!\!\!\!\!\!${}^{1}$Institute of Mathematical Sciences, Ewha Woman's University,
52 Ewhayeodae-gil, Seodaemun-gu, Seoul  03760, KOREA}\\
${}^{2}$Department of Physics, Sogang University, 35 Baekbeom-ro, Mapo-gu,  Seoul  04107, KOREA\\
${}^{3}$Department of Physics, McGill University, Montr\'{e}al, QC, H3A 2T8, CANADA\\
\mbox{\!\!\!\!\!\!\!\!\!\!\!\!\!${}^{4}$Center for Gravitational Physics, 
Yukawa Institute for Theoretical Physics, Kyoto University,
 606-8502, Kyoto, JAPAN}\\
\mbox{${}^{5}$Kavli  Institute  for  the  Physics  and  Mathematics  of  the  Universe  (WPI)}\\
\mbox{\!\!\!\!\!\!\!\!\!\!\!\!\!\!\!\!The  University  of  Tokyo  Institutes  for  Advanced  Study, The  University  of  Tokyo,  Kashiwa,  Chiba  277-8583,  JAPAN}\\
~\\
\texttt{sangus@ewha.ac.kr\qquad 
khcho23@sogang.ac.kr\qquad
guilherme.franzmann@mail.mcgill.ca\qquad
shinji.mukohyama@yukawa.kyoto-u.ac.jp\qquad park@sogang.ac.kr}\\
~\\
\end{center}
\begin{abstract} 
\vspace{-12pt}
\centering\begin{minipage}{\dimexpr\paperwidth-6.7cm}
\noindent  
{In string theory the closed-string massless NS-NS sector forms a multiplet of $\mathbf{O}(D,D)$ symmetry.  This suggests a specific modification to General Relativity in which the entire NS-NS sector is promoted to stringy graviton fields.  Imposing off-shell  $\mathbf{O}(D,D)$ symmetry  fixes  the correct couplings  to other matter fields and the  Einstein field equations are enriched  to comprise $D^{2}+1$ components, dubbed recently as the Einstein Double Field Equations.    Here we explore the cosmological  implications of this framework.   We derive  the most general homogeneous and isotropic ansatzes for both stringy graviton fields and the $\mathbf{O}(D,D)$-covariant energy-momentum tensor. Crucially,  the former admits  space-filling  magnetic $H$-flux. Substituting them into the Einstein Double Field Equations, we obtain the $\mathbf{O}(D,D)$ completion of the  Friedmann equations along with a generalized continuity equation. 
We  discuss how solutions in this framework may be characterized by two equation-of-state parameters,  $w$ and $\lambda$, where the latter characterizes the relative intensities of scalar and tensor forces. 
When $\lambda+3w=1$, the dilaton remains constant throughout the cosmological evolution, and one  recovers  the standard Friedmann equations for generic matter content (\textit{i.e.} for any $w$).
We further point out that, in contrast to General Relativity, neither an $\mathbf{O}(D,D)$-symmetric cosmological constant nor a scalar field with positive energy density gives rise to a de Sitter solution. }
\end{minipage}

\end{abstract} 
{\small
\begin{flushright}
\textit{Preprint}: YITP-19-40, IPMU19-0074
\end{flushright}}
\thispagestyle{empty}
\end{titlepage}
\newpage

\tableofcontents 




%

\section{Introduction}
Despite its many successes, General Relativity (GR)  faces several  well-known shortcomings  when applied to cosmology.  In order to explain the large-scale dynamics of the universe, one needs to introduce dark matter and dark energy.  Furthermore, solving the horizon and flatness problems requires new dynamics, such as inflation or bouncing cosmologies, involving additional degrees of freedom which may come into play near the strong-coupling regime at which GR breaks down.

In order to make quantitative predictions about the very early universe, we  need to invoke a consistent theory of quantum gravity.  String theory is currently the best-developed candidate, and its effects on cosmology may be studied in the low-energy effective supergravity (SUGRA) limit at weak coupling~\cite{Gasperini:2002bn}.
However, string theory does not predict GR exactly.   In GR the spacetime metric, $g_{\mu\nu}$, is the only gravitational field.  On the other hand, string theory predicts its own gravity,  or Stringy Gravity, of which the fundamental  fields consist of   the massless modes of the closed string (or  superstring NS-NS sector): the metric, $g_{\mu\nu}$, the Kalb--Ramond (NS-NS) two-form, $B_{\mu\nu}$, and the dilaton, $\phi$.  Traditionally, in the search for superstring vacua, one treats the dilaton and $B$-field as moduli which should be dynamically stabilized by some mechanism \cite{Damour:1994ya}.\footnote{However, the $\ODD$ symmetry fixes the minimal coupling of the closed-string massless sector to a point particle, which in string frame results in the usual  relativistic particle action prescribing geodesic motion. Thus,  for a point particle the Equivalence Principle is still valid in string frame,  whether  the dilaton is frozen or not~\cite{Ko:2016dxa}.}
Assuming this has been done, one  often then performs a Weyl transformation on the metric, bringing it from the original `string frame' to the `Einstein frame': $g_{\mu\nu}^{\textrm E} = g_{\mu\nu}\exp(-2\phi_{0})$, where $\phi_{0}$ is the dilaton vacuum expectation value in four-dimensional spacetime.

{More generally, one can consider the other closed-string modes, the scalar dilaton $\phi$ and the Kalb--Ramond two-form potential $B_{\mu\nu}$,  as dynamical degrees of freedom in addition to the spacetime metric. The inclusion of the dilaton  leads to the well-studied SUGRA cosmology, which has a long and rich history \cite{Gasperini:2002bn,Tseytlin:1991xk,Brustein:1997ny,Gasperini:2007ar,Gasperini:2004ss,Gasperini:2007zz,Meissner:1991zj,Meissner:1991ge,Gasperini:1991ak}. However, it is not clear within the framework of SUGRA how one should couple the matter sector to the gravitational sector, especially regarding the dilaton field and the $B$-field. Indeed, most prior work on this topic considers a minimal coupling to the metric and no coupling to the  dilaton  (for an exception, see \cite{Gasperini:2007ar,Gasperini:2004ss}\footnote{{ Note that the authors of  \cite{Gasperini:2007ar,Gasperini:2004ss} considered spatial $\mathbf{O}(d,d)$ symmetry  in order to fix the matter-dilaton coupling, an on-shell symmetry present only at the background level, whereas the  $\mathbf{O}(D,D)$ symmetry of Double Field Theory is off-shell, valid for any background.}}) nor to the $B$-field (\textit{c.f.~}dual scalar axion~\cite{Svrcek:2006yi}). It would be desirable to find a principle which can constrain the allowed coupling on a more fundamental level.}

{To this end, we recall that symmetry has been the guiding principle of modern physics, dictating which interactions are allowed.  In SUGRA the  whole set of closed-string massless NS-NS fields $\{g_{\mu\nu},B_{\mu\nu},\phi\}$ transform into each other under $\ODD$ rotations, where $D$ is the number of spacetime dimensions.}  However, $\ODD$ is not a symmetry in the usual sense, as a particular choice of $\{g_{\mu\nu},B_{\mu\nu},\phi\}$, \textit{i.e.} a particular \textit{background}, typically preserves at most only a residual discrete subgroup of $\ODD$, known as T-duality. A well-known example of T-duality is the small-large duality $R \leftrightarrow 1/R$, which heuristically leads to the exchange of `momentum' and `winding' modes of the closed  string~\hbox{\cite{Buscher:1987sk,Buscher:1987qj}.}  In recent years {$\ODD$ has been elevated to a manifest symmetry in the formalism of Double Field Theory (DFT)~\cite{Siegel:1993xq,Siegel:1993th,Hull:2009mi,Hull:2009zb,Hohm:2010jy,Hohm:2010pp} (as well as doubled string worldsheet  actions~\cite{Duff:1989tf,Tseytlin:1990nb,Tseytlin:1990va,
Hull:2004in,Hull:2006va}), } in which \hbox{$D+D$}~coordinates are used to describe $D$-dimensional physics, with  actual physical points identified as gauge orbits in the doubled coordinate system~\cite{Park:2013mpa,Lee:2013hma}.  {This framework encompasses not only different SUGRA backgrounds such as Type IIA and IIB SUGRA, but also includes} non-Riemannian gravities such as Newton--Cartan, ultra-relativistic Carroll, \textit{etc.}, all of which are unified as different backgrounds of a single theory~\cite{Morand:2017fnv}. 

It is therefore natural to expect $\ODD$, when interpreted as an exact symmetry of the full non-perturbative  string theory, to constrain the form of allowed interactions.  With this in mind, in this work we assume the $\ODD$ symmetry as a 
first principle.  That is to say, although the initial motivation of DFT was to reformulate SUGRA in an $\ODD$ manifest manner, one may well postulate  this `$\ODD$ Principle' as a starting point, \textit{a priori} independent of supersymmetry.  In this spirit,  DFT has recently been extended to also incorporate matter content --- not only conventional string modes such as the Ramond-Ramond \cite{Rocen:2010bk,Hohm:2011zr,Jeon:2012kd} or R-NS sectors \cite{Jeon:2011vx,Jeon:2011sq,Jeon:2012hp}, but also the Standard Model~\cite{Choi:2015bga}~\textit{c.f.~}\cite{Jeon:2011kp} --- in a consistent  manner which preserves the $\ODD$ symmetry.\footnote{If the Standard Model is to be realized within string theory, one may wonder if the $\ODD$ symmetry is preserved in Nature. In fact, \cite{Choi:2015bga} shows that the pure Standard Model 
without any extra physical degrees of freedom can be coupled to DFT in a manner compatible with the $\ODD$ Principle.}  The interaction between DFT and (generic) matter is then described via the  
so-called Einstein Double Field Equations (EDFEs)~\cite{Angus:2018mep}  (see also \cite{Park:2019hbc} for a  short summary), which 
accommodate the gravitational sector together with matter while preserving invariance under the $\ODD$ symmetry.  This `Stringy Gravity' thus represents the \textit{$\ODD$-completion} of General Relativity and generalizes SUGRA cosmology.  Earlier   discussions   on  the cosmological  implications  of the  $\ODD$ symmetry include  \cite{Wu:2013sha,Wu:2013ixa,Ma:2014ala} while other attempts to incorporate matter have also been considered in  \cite{Brandenberger:2017umf,Brandenberger:2018xwl,Brandenberger:2018bdc,Bernardo:2019pnq}.

With the EDFEs as our starting point, in this work we derive the generalized Friedmann equations  which follow from this overarching $\ODD$-symmetric framework.  Hereafter, we refer to them as the \emph{$\ODD$-complete Friedmann Equations} (OFEs).    As will be shown, imposing the $\ODD$ Principle results in modifications to the conventional SUGRA equations for Riemannian backgrounds.  Perhaps surprisingly, one finds that
whenever the dilaton is kept constant, whether dynamically or through some appropriate coupling to the matter sector,
the standard Friedmann equations are recovered in the presence of any matter sources, not just for a radiation-dominated universe as is the case in the usual string cosmology \cite{Tseytlin:1991xk}.

{We stress that by requiring interactions to be consistent with the underlying off-shell $\ODD$   symmetry, we fix uniquely the couplings of $\{g_{\mu\nu},B_{\mu\nu},\phi\}$ to matter.  Hence, while this `top-down' approach reproduces many known results in string cosmology and recasts them into a more general framework, the $\ODD$ Principle may also be interpreted as a restricting criterion for model-building.}

The organization of the present paper is as follows.  In  section~\ref{SECReview}  we review briefly the Einstein Double Field Equations~\cite{Angus:2018mep,Park:2019hbc}. In 
section~\ref{SECOFE} we derive our main result, the $\ODD$ complete Friedman Equations with $D=4$.  In particular, we introduce  \textit{two} equation-of-state parameters, $w$ (usual) and a new parameter $\lambda$ { which measures the density ratio between scalar (dilaton) and tensor (metric) gravitational forces \cite{Gasperini:2007ar}.}  In the following subsections we apply some DFT results from \cite{Angus:2018mep} to cosmology: in section~\ref{sec:emt} we summarize the stringy energy-momentum tensors of various types of matter; and in section~\ref{sec:econd} we discuss energy conditions.
Appendix~\ref{APPENDIXKilling} contains a derivation of the  most general  cosmological, \textit{i.e.~}homogeneous and isotropic,  form of the stringy energy-momentum tensor within the framework of Double Field Theory. 
Section~\ref{SECSolutions}  discusses various solutions, such as a (generalized) perfect fluid, scalar field, and radiation.
 
In standard cosmology there are several key scenarios in which the evolution of the universe approaches a de Sitter spacetime.  These include the late-time $\Lambda$-dominated expansion and the hypothesized period of inflation in the early universe.  However, in recent years the question of whether or not de Sitter solutions can be realized in a consistent theory of quantum gravity, such as string theory, has been widely debated (see, for example, \cite{Obied:2018sgi,Agrawal:2018own,Andriot:2018wzk,Garg:2018reu,Conlon:2018eyr,Kachru:2018aqn,Akrami:2018ylq,Murayama:2018lie,Hamaguchi:2018vtv,Ooguri:2018wrx,Hebecker:2018vxz,Danielsson:2018ztv,Kobakhidze:2019ppv,Andriot:2019wrs,Dvali:2018jhn}).  Hence in section \ref{SECdeS} we investigate in detail the possibility of realizing de Sitter solutions in $\ODD$-symmetric cosmology, and find that it appears to require solutions with negative energy density, a violation of the weak energy condition.
We conclude with comments in section~\ref{SECCON}.


\section{The $\ODD$ paradigm: review of the Einstein Double Field Equations\label{SECReview}}

In General Relativity (GR) the metric, which sets the local geometry, is the only field responsible for gravitational phenomena. All other fields are classified as additional matter and couple unambiguously to the geometry via a minimal coupling, \textit{i.e.}~promoting ordinary derivatives to covariant ones and generalizing volume elements. This procedure ensures covariance under both diffeomorphisms and, with the vielbein formalism, local Lorentz symmetry. 

The situation is more involved in supergravity, in which the (bosonic)  gravitational sector includes three different fields: the metric $g_{\mu\nu}$, the $B$-field $B_{\mu\nu}$, and the dilaton $\phi$. Together with the symmetries from GR, SUGRA is also invariant under gauge transformations of the $B$-field. 
Although both setups are very similar, in SUGRA not all of the fields responsible for gravity are also geometric: that role is exclusively reserved for the metric. Only since the introduction of DFT~\cite{Siegel:1993xq,Siegel:1993th,Hull:2009mi,Hull:2009zb,Hohm:2010jy,Hohm:2010pp} has it been possible to consider all gravitational fields on the same footing, as being responsible for defining geometry. Note that the DFT geometry is still not fully understood  in the mathematical literature,  and although it recovers Riemannian backgrounds under certain conditions, it is certainly much more general, see \textit{e.g.~} \cite{Hitchin:2004ut,Gualtieri:2003dx,
Hitchin:2010qz,Coimbra:2011nw,Coimbra:2012yy,Vaisman:2012ke,Berman:2013uda,Garcia-Fernandez:2013gja,Cederwall:2014kxa,Cederwall:2016ukd,Deser:2016qkw,Sakamoto:2017cpu,Morand:2017fnv,Cederwall:2017fjm,Freidel:2018tkj,Chatzistavrakidis:2019huz}.  The geometrical framework of DFT  allows us to define a generalization of the scalar and Ricci curvatures, $\So$ and $P_{A}{}^{C}\brP_{B}{}^{D}S_{CD}$~\cite{Jeon:2011cn} (\textit{c.f.}~\cite{Jeon:2010rw,Hohm:2011si}), respectively,  both of which may reduce to the usual definitions with a trivial dilaton and $B$-field. The natural next step is to include matter content into such a framework: this was accomplished in  \cite{Choi:2015bga} and the resulting interactions between gravity and matter are described by the \hbox{Einstein Double Field Equations (EDFEs)}~\cite{Angus:2018mep,Park:2019hbc},
\begin{equation}
    G_{AB} = 8\pi G T_{AB} \, , \label{GenEinstein}
\end{equation}
where the above indices are charged under the $\ODD$ symmetry group. The left-hand side corresponds to the string-theoretic extension of the Einstein tensor which is conserved off-shell~\cite{Park:2015bza},\footnote{$\ODD$ indices are lowered and raised by $\cJ_{AB}$ and its inverse.  For more details on various aspects of DFT, see Appendix \ref{APPENDIXKilling}.}
\be
\ba{ll}
\dis{G_{AB}=4P_{[A}{}^{C}\brP_{B]}{}^{D}S_{CD}-\frac{1}{2}\cJ_{AB}\So\,,
}\qquad&\qquad
\cD_{A}G^{AB}=0\,,
\ea
\ee
while the right-hand side is the corresponding generalization of the energy-momentum tensor, 
which is conserved on-shell~\cite{Angus:2018mep}, 
\be \label{stringEMT}
\ba{ll}\dis{
T^{AB}=e^{2d}\left(8
\brP^{[A}{}_{C}P^{B]}{}_{D}\frac{\delta\cL_{{\rm{matter}}}}{\delta\cH_{CD}}-\frac{1}{2}\cJ^{AB}\frac{\delta\cL_{{\rm{matter}}}}{\delta d}\right)\,},\qquad&\quad\cD_{A}T^{AB}=0\,.
\ea
\ee
These objects respect all symmetries of DFT.

Our starting point for this paper is to consider purely Riemannian backgrounds in the above equation (\textit{c.f.~}non-Riemannian ones~\cite{Morand:2017fnv}). In such cases, \eqref{GenEinstein} reduces to
\begin{eqnarray}
R_{\mu\nu}+2\trd_{\mu}(\partial_{\nu}\phi)-\quarter H_{\mu\rho\sigma}H_{\nu}{}^{\rho\sigma}\nonumber
&=&8\pi G\EK_{(\mu\nu)}\,,\\
\trd^{\rho}\!\left(e^{-2\phi}H_{\rho\mu\nu}\right)&=&16\pi Ge^{-2\phi}\EK_{[\mu\nu]}\,,\label{EDFE}\\
R+4\Box\phi-4\partial_{\mu}\phi\partial^{\mu}\phi-\textstyle{\frac{1}{12}}H_{\lambda\mu\nu}H^{\lambda\mu\nu}&=&8\pi G\To\,, \nonumber
\end{eqnarray}
where $R$ and $R_{\mu\nu}$ stand for the usual Ricci scalar and tensor  in string frame, respectively, while $H_{\mu\nu\rho}$ is the field strength of the $B$-field. The   skew-symmetric   $K_{[\mu\nu]}$ and the symmetric $K_{(\mu\nu)}$    can be understood (up to equations of motion) as the matter sources  for the $B$-field and the traceless part of the metric, respectively, while $\To$ corresponds to the matter sourcing 
the trace of the metric and the dilaton,
\be
\To:=e^{2d}\times\frac{\delta \cL_{{\rm{matter}}}}{\delta d}\,.
\ee
The terms on the right-hand side of the latter two equations   in (\ref{EDFE}) are absent in  conventional SUGRA (though the dilaton coupling has been considered in the context of spatial O$(d,d)$ in \cite{Gasperini:2007ar,Gasperini:2004ss}), and their inevitable inclusion above characterizes a modification, or conceptual generalization, of standard 
SUGRA cosmology in terms of how the matter sources are coupled to the stringy gravity sector.  The inclusion of these terms puts
all  gravitational fields on an equal footing, and crucially is a direct result of imposing thoroughly the $\ODD$ symmetry on the DFT action coupled to matter.

Together with the above equations, there is also an on-shell conservation law for the stringy energy-momentum tensor, arising from `doubled' general covariance.  This is also consistent with the fact that the stringy Einstein curvature tensor is, by construction, covariantly conserved off-shell \cite{Park:2015bza}.  On Riemannian backgrounds this conservation law reduces to the equations
\be
\na^{\mu}\EK_{(\mu\nu)}-2\partial^{\mu}\phi\,\EK_{(\mu\nu)}+\half H_{\nu}{}^{\lambda\mu}\EK_{[\lambda\mu]}-\half\partial_{\nu}\To=0\,,
\qquad\na^{\mu}\!\left(e^{-2\phi}\EK_{[\mu\nu]}\right)=0\,.
\label{ConLaw}
\ee

The above equations may be derived from the spacetime action 
\be
\dis{\int d^4 x \sqrt{-g}e^{-2\phi}\Big[\,\textstyle{\frac{1}{16\pi G}}\left( R+4\Box\phi-4\partial_{\mu}\phi\partial^{\mu}\phi-\textstyle{\frac{1}{12}}H_{\lambda\mu\nu}H^{\lambda\mu\nu} \right) +L_{\rm{matter}}\,\Big]\,}
\label{SGaction}.
\ee
This differs from the usual SUGRA cosmology action by the fact that the $\ODD$-invariant measure, $e^{-2d}\equiv\sqrt{-g}e^{-2\phi}$, couples to the entire matter Lagrangian as $\sqrt{-g}e^{-2\phi}L_{\rm{matter}}\equiv\cL_{\rm{matter}}$ (where $\cL_{\rm{matter}}$ is the Lagrangian density of \eqref{stringEMT}) and hence matter is coupled not only to the metric but also to the dilaton and $B$-field (through the covariant derivatives).  Also note that the metric $g_{\mu\nu}$ above is the string (Jordan)-frame metric rather than the Einstein-frame metric.


\section{$\ODD$ completion of the  Friedmann equations \label{SECOFE}}

In this section we obtain the most general ansatz for a $D=4$ homogeneous and isotropic cosmological background.  This allows us to write down the resulting  $\ODD$ completion of the   Friedmann equations.

The DFT-Killing equations for Riemannian backgrounds are given by \cite{Angus:2018mep} (see Appendix \ref{APPENDIXKilling})
\be
\cL_{\zeta_{a}}g_{\mu\nu}=0\,,\quad\qquad
\cL_{\zeta_{a}}B_{\mu\nu}
+\partial_{\mu}\tilde{\zeta}_{a\nu}-\partial_{\nu}\tilde{\zeta}_{a\mu}=0\,,\quad\qquad\cL_{\zeta_{a}}\phi=0\,,
\ee
where $\zeta_{a}$ are ordinary GR Killing vectors, while the $\tilde{\zeta}_{a}$ are corresponding one-forms required to complete the parametrization of DFT isometries.  In order to study cosmology we should consider homogeneous and isotropic backgrounds, in which case these DFT-Killing vectors will correspond to spatial rotations and translations.  In such cases the most general solution is given by\footnote{Here $B_{\scriptscriptstyle{(2)}}$ is defined up to a gauge term; the resulting $H$-flux \eqref{isoH} is gauge invariant.}
\be
\ba{cll}
\rd s^{2}&=&-N^2(t)\rd t^{2} +a^2(t)\left[ \frac{1}{1-kr^{2}} dr^2 +r^2 d \Omega^{2}\right]\,,\\
B_{\scriptscriptstyle{(2)}}&=&\frac{h r^{2}}{\sqrt{1-kr^{2}}}\cos\vartheta\,\rd r\wedge\rd\varphi \,,\\
\phi& =& \phi(t)\,, \label{CosmoAnsatz}
\ea
\ee
where  $\rd\Omega^{2}=\rd\vartheta^{2}+\sin^{2}\vartheta\rd\varphi^{2}$, $B_{\scriptscriptstyle{(2)}}=\half B_{\mu\nu}\rd x^{\mu}\wedge\rd x^{\nu}$, $k$ corresponds to the spatial curvature, and $h$ is a constant corresponding to the magnetic $H$-flux
\begin{equation}\textstyle{
H_{\scriptscriptstyle{(3)}}=\frac{h r^{2}}{\sqrt{1-kr^{2}}}\sin\vartheta\,\rd r\wedge\rd\vartheta\wedge\rd\varphi \,.}  \label{isoH}
\end{equation}
We emphasize that non-trivial $H$-flux is compatible with the cosmological principle only for $D=4$.  Note that the lapse function $N(t)$ simply fixes the $t$-coordinate gauge, so at leading order the solutions are fully characterized by the functions $a(t)$ and $\phi(t)$, as well as the parameters $h$ and $k$.  Moreover, we must impose the same symmetry conditions on the matter sector, resulting in the Killing equations\footnote{Note that the equations of motion \eqref{EDFE} imply that for homogeneous and isotropic solutions \eqref{CosmoAnsatz}, the antisymmetric part of $\EK_{\mu\nu}$ must vanish.}
\be
\cL_{\zeta_{a}}\EK_{\mu\nu}=0\,, \qquad\cL_{\zeta_{a}}\To=0\,.
\ee
The latter implies that $\To(t)$ is a time-dependent function, while the former implies that $K_{\mu\nu}$ is diagonal and spatially homogeneous,
\be
\EK^{\mu}{}_{\nu}=\left(\ba{cc}
~\EK^{t}{}_{t}(t)~&~0~\\
~0~&~\EK^{r}{}_{r}(t)\delta^{i}{}_{j}~
\ea\right)\,, \label{CosmoEMT}
\ee
where $K^{t}{}_{t}(t)$ and $K^{r}{}_{r}(t)$ are time-dependent functions, and $K^{1}{}_{1} = K^{2}{}_{2} = K^{3}{}_{3} \equiv K^{r}{}_{r}(t)$.\footnote{Since the above ansatz has been written for $D=4$, we are not considering critical strings. However, it can easily be generalized to arbitrary dimensions   (but note that the possibility of an isotropic 3-form $H$-flux \eqref{isoH} is unique to three spatial dimensions).}

Applying the ansatz \eqref{CosmoAnsatz} and \eqref{CosmoEMT} to \eqref{EDFE}, we obtain the $\ODD$-complete  Friedmann equations,
\begin{align}
  &8\pi G  (K^t{}_t + 3 K^r{}_r - \To) N^2
  = -\frac{h^2 N^2}{a^6} + 6H\phi' - \frac{2 N'\phi'}{N} - 4(\phi')^2 + 2 \phi'' \,,\nonumber \\
  & 8 \pi G  K^t{}_t N^2 = -\frac{3 HN'}{N} + \frac{2 N'\phi'}{N} + \frac{3a''}{a} - 2\phi'' \,, \label{gfe} \\
  & 8 \pi G  K^r{}_r N^2 = -\frac{h^2 N^2}{2 a^6} + \frac{2 k N^2}{a^2} + 2H^2 -\frac{H N'}{N} - 2 H\phi' + \frac{a''}{a}\,, \nonumber
\end{align}
where $H\equiv a'/(Na)$ and the prime denotes differentiation with respect to the time coordinate of \eqref{CosmoAnsatz}.  Furthermore, applying the ansatz to \eqref{ConLaw} yields one non-trivial conservation equation,
\begin{align}
  \frac{\rd~}{\rd t} \Big( K^t{}_t -\frac{1}{2} \To \Big) = 3NH (K^r{}_r - K^t{}_t) + 2 \phi' K^t{}_t\,. \label{conservation1} 
\end{align}

In order to make contact with known physics we must rewrite these equations in terms of standard physical quantities such as energy density and pressure.  Our basic assumption is that standard FLRW cosmology should be recovered from \eqref{gfe} and \eqref{conservation1} in the case where the dilaton is constant, $\phi' = \phi'' = 0$, and the $H$-flux vanishes, $h = 0$.  One might also be tempted to set $\To = 0$, however this cannot in general be the case, as seen, for example, from the first equation of \eqref{gfe}: instead we find $\To = K^{\mu}{}_{\mu}$ in this limit. Thus we should find a definition of energy density and pressure in terms of $K^{\mu}{}_{\nu}$ and $\To$. We propose 
\begin{align}
  &\rho :=\left(  - K^t{}_t + \frac{1}{2} \To \right) e^{-2\phi}\, , \qquad p := \left(K^r{}_r -\frac{1}{2} \To  \right) e^{-2\phi} \,. \label{setrhop}
\end{align}
As we will see shortly, this definition reduces \eqref{gfe} to the standard Friedmann equations in the limit of constant dilaton and vanishing $H$-flux.  It is further justified from writing the Hamiltonian of (\ref{SGaction}) in a cosmological background, from which one can easily see that the corresponding energy density should be given by the above formula.  Note that the $e^{-2\phi}$ factors arise as a direct consequence of the same overall factor appearing in the $\ODD$-invariant matter action~\eqref{SGaction}.  As shown below, it is through this identification that the SUGRA cosmological equations can also be recovered.

Using (\ref{setrhop}) and rearranging (\ref{gfe}), we obtain our primary result.
\vspace{-0.2cm}
\begin{framed}
\noindent
The $\ODD$-complete Friedmann Equations (OFEs) are\
\begin{eqnarray}
\frac{8\pi G}{3}\rho e^{2\phi} + \frac{h^2}{12a^6} &=& H^{2} - 2\left(\frac{\phi^{\prime}}{N}\right)H + \frac{2}{3}\left(\frac{\phi^{\prime}}{N}\right)^2 +  \frac{k}{a^2} \, , \label{OFE1}
\\ \frac{4\pi G}{3} (\rho+3p)e^{2\phi} + \frac{h^2}{6a^6} &=& - H^{2} - \frac{H^{\prime}}{N} + \left(\frac{\phi^{\prime}}{N}\right)H - \frac{2}{3}\left(\frac{\phi^{\prime}}{N}\right)^2 + \frac{1}{N}\left(\frac{\phi^{\prime}}{N}\right)' \, , \label{OFE2}
\\ \frac{8\pi G}{3}\left(\rho e^{2\phi} - \frac{1}{2}\To\right) &=& - H^{2} - \frac{H^{\prime}}{N} + \frac{2}{3N}\left(\frac{\phi^{\prime}}{N}\right)' \, , \label{OFE3}
\end{eqnarray}
which imply  the conservation equation (c.f. \cite{Brustein:1997ny})
\begin{align}
  \rho^{\prime}+3NH (\rho +p) + \phi^{\prime} \To e^{-2\phi} =0 \,.\label{Conservation}
\end{align}
\vspace{-0.8cm}
\end{framed}
\noindent Specifically when $h=k=0$, the cosmological ansatz and the OFEs are preserved under the  entire spatial T-duality, given in Table~\ref{Trule} below.\footnote{Although the T-duality relation in Table~\ref{Trule} holds for the special case $h = k = 0$ only, we nevertheless describe \eqref{OFE1}, \eqref{OFE2}, and \eqref{OFE3} as `$\ODD$-complete' for any $h$ and $k$, since they are the cosmological equations of motion for $\ODD$-symmetric Lagrangians.  Namely, we impose the $\ODD$ Principle on the underlying Lagrangian description, but do not necessarily require the solutions to be $\ODD$-symmetric.}
\begin{table}[H]
\quad\!\begin{tabular}{|c|c|c|c|c|c|c|c|c|c|}
\hline
Before~&$~N$~& ~$a$~& ~$H$~& $\phi$ &  $~\rho$~&  $p$&  ~$\To$~&  $~K^{t}{}{}_{t}$~&  ~$K^{r}{}_{r}~$\\
\hline
After~&$~N~$& $a^{-1}$& $~-H~$& $~\phi-3\ln a~$& $~a^{6}\rho~$& $~-a^{6}\left(p+\To e^{-2\phi}\right)~$& 
$~\To~$& $~K^{t}{}{}_{t}~$& $~-K^{r}{}_{r}~$\\
\hline
\end{tabular}
\caption{Cosmological T-duality transformations $(h=k=0)$.}
\label{Trule}
\end{table}
\noindent 

From (\ref{OFE1}) we may solve for the time derivative of the dilaton,
\be
\frac{2\phi^{\prime}}{N}=3H\pm\sqrt{3H^{2}+16\pi G\rho e^{2\phi}-\frac{6k}{a^{2}}+\frac{h^{2}}{2a^{6}}}\,.
\label{phidot}
\ee
Substituting this into either (\ref{OFE2}) or (\ref{OFE3}), up to the conservation relation~(\ref{Conservation}),  we obtain an expression for the  time evolution of $H$,
\be
\frac{H^{\prime}}{N}=8\pi G\left(pe^{2\phi}+\frac{1}{2}\To\right)
-\frac{2k}{a^{2}}+\frac{h^{2}}{2a^{6}}\pm H\sqrt{3H^{2}+16\pi G\rho e^{2\phi}-\frac{6k}{a^{2}}+\frac{h^{2}}{2a^{6}}}\,.
\label{Hdot}
\ee
In fact, the OFEs,~(\ref{OFE1}), (\ref{OFE2}), and (\ref{OFE3}), are equivalent to (\ref{Conservation}), (\ref{phidot}), and (\ref{Hdot}).
It also is worthwhile to rearrange \eqref{OFE1} and \eqref{OFE2} to obtain
\begin{eqnarray}
4\pi Gp e^{2\phi} + \frac{h^2}{8a^6} &=& - \frac{3}{2}H^{2} + \frac{H'}{N} + 2\left(\frac{\phi'}{N}\right)H - \left(\frac{\phi'}{N}\right)^2 + \frac{1}{N}\left(\frac{\phi'}{N}\right)' - \frac{k}{2a^2}\,,
\label{OFEp}\\ 
4\pi G(\rho + p)e^{2\phi}+ \frac{h^{2}}{4a^{6}} &=&   - \frac{H'}{N} - \left(\frac{\phi'}{N}\right)H + \frac{1}{N}\left(\frac{\phi'}{N}\right)'+\frac{k}{a^{2}} \, .
\label{extra}
\end{eqnarray}
Then all the terms appearing on the left-hand sides of  \eqref{OFE1}, \eqref{OFE2}, \eqref{OFE3},  \eqref{OFEp}, and \eqref{extra} correspond precisely to the quantities appearing in various energy conditions, as will be discussed further in section \ref{sec:econd}.

These OFEs represent conceptually a cosmology derived first and foremost from the $\ODD$ Principle, which may also extend beyond SUGRA.
We also re-emphasize that here 
the matter sector is coupled covariantly to the dilaton as well as the string-frame metric, which is again a direct consequence of extending the $\ODD$ symmetry of the DFT gravitational sector to a generic matter sector.  In practice the origin of matter in $D$ dimensions can be diverse: while it may arise directly from the massless sectors of ten-dimensional superstring theories, in lower-dimensional effective field theories (which may break supersymmetry) any coupling which is not forbidden should in principle appear.  The equations \eqref{OFE1}--\eqref{OFE3} may then characterize the cosmological evolution of any matter contributions which respect the $\ODD$ Principle (but not necessarily supersymmetry). 

It is known that in SUGRA, when the dilaton is stabilized (in the absence of spatial curvature and with trivial $B$-field) one recovers a radiation-dominated universe, with a linear barotropic equation of state given by $w=1/3$ in $D=4$ \cite{Tseytlin:1991xk}. In the present case, for vanishing $H$-flux and constant dilaton, for which we choose $\phi = 0$ without loss of generality, the OFEs reduce to
\begin{align}
   &H^{2} = \frac{8 \pi G}{3}\rho - \frac{k}{a^{2}} \nonumber \,, \\
   & \frac{\ddot{a}}{a} = -\frac{4\pi G}{3}(\rho + 3p) \,, \label{GRlimit} \\ 
   & \To = \rho-3p \nonumber \,,
\end{align}
where we have chosen cosmic gauge ($N=1$), with the dot `$\dot{~}$' denoting differentiation with respect to cosmic time.
The first two equations are the standard Friedmann equations for generic matter content, while the final equation fixes $\To$ in terms of density and pressure.  Therefore, the low-energy DFT limit with a stabilized dilaton is consistent with introducing any type of matter (including extended objects such as D-branes and NS-branes \cite{Berkeley:2014nza,Berman:2014jsa,Blair:2017hhy,Blair:2018lbh,Blair:2019rrj}), as opposed to only radiation. This is surprising and hints towards the fact that the current framework might be the natural extension of GR to higher energies. 

Looking to the last equation above, it is also clear that in the limit of vanishing $\To$ one recovers standard SUGRA cosmology, for which the equation of state corresponds to radiation if the dilaton is stabilized. With that in mind, we may define the ratios:
\begin{equation}
w := \frac{p}{\rho} \, ; \qquad \lambda := \frac{\To e^{-2\phi}}{\rho} \, . \label{lambda}
\end{equation}
In general these are not necessarily constant, but may be time-dependent functions.  However, in the case where they are constant,\footnote{This generalizes a linear barotropic equation of state, with $\rho/p = 1/w$ and $\To e^{-2\phi}/p = \lambda/w$.} $w$ is the conventional parameter corresponding to the effective pressure of the generalized fluid, while the new parameter, $\lambda$, measures the \emph{density rate} at which the matter is coupled to the dilaton in comparison with the coupling to the metric \cite{Gasperini:2007ar}. Since the dilaton is now also part of the gravity sector, naturally one can understand dilaton-induced interactions as an additional component of the gravitational interactions of matter. 
Note that for a general barotropic fluid the quantity $\lambda$ should be provided \textit{a priori}, so one can think of it as providing a generalized equation of state, in conjunction with $w$.  In general, we expect that different types of matter satisfying the $\ODD$ Principle will yield various values of $w$ and $\lambda$, since in our framework $\rho$, $p$, and $\To$ are independent \textit{a priori} (at least conceptually) due to the dilatonic coupling to matter (for concrete examples, see section 2.2 of~\cite{Angus:2018mep}).
Thus a perfect fluid with a linear equation of state in an $\ODD$-covariant cosmological background would be characterized by two parameters, $w$ and $\lambda$. We will discuss this further and present some examples in section~\ref{SECSolutions}. 

The appearance of $\To$ is the main feature that distinguishes the OFEs from the standard SUGRA cosmological backgrounds studied in the literature {(see~\cite{Gasperini:2002bn} for a review)}. Thus in the next subsection we summarize how the stringy energy-momentum tensor, in particular $\To$, is evaluated for different types of matter content.  Note that in all cases where the dilaton is minimally coupled (\emph{i.e.} via the DFT volume element) to a dilaton-independent spacetime Lagrangian $L_{\textrm{matter}}$, we find $\To = -2L_{\textrm{ matter}}$.  Hence in such cases, vanishing $\To$ simply corresponds to the Lagrangian vanishing on-shell.

\subsection{Examples of stringy energy-momentum tensors in cosmology}
\label{sec:emt}

In \cite{Angus:2018mep} many different examples of matter content were considered, with the stringy energy-momentum tensor components $K_{\mu\nu}$ and $\To$ computed for each.  Here we collect and summarize these results, and comment on their respective cosmological implications.

\begin{itemize}
    \item \textbf{Cosmological constant:} The DFT cosmological constant simply couples minimally to the DFT volume element, which crucially includes the dilaton~\cite{Jeon:2011cn}.  Varying such a term in the action gives a non-vanishing contribution only for $\To$, such that $K_{\mu\nu} = 0$ and $\To=\frac{1}{4\pi G}\Lambda_{\DFT}$.
    
    \item \textbf{Scalar field:} The canonical Lagrangian for a scalar field $\Phi$ in DFT also couples minimally to the dilaton, and further couples linearly to the (inverse) string-frame metric, giving $K_{\mu\nu} = \p_{\mu}\Phi\p_{\nu}\Phi$ and $\To=-2L_{\Phi}$.  In particular, we will see later in \eqref{wlscalar} that this implies $\lambda = -2w$.  The presence of non-zero $\To$ in the OFEs should have intriguing consequences for the general dynamics of a scalar field in cosmological $\ODD$-symmetric backgrounds. In particular, inflationary models in the context of supergravity should be revisited in future work. 
    
    \item \textbf{Fermionic fields:} The fermionic Lagrangian is proportional to its equation of motion, thus minimal dilaton coupling implies that $\To$ vanishes on-shell.  Furthermore, it turns out that in this case we may change variables to a spinor density whose Lagrangian decouples completely from the dilaton, giving $\To = 0$ even off-shell.  However, in general $K_{\mu\nu} = -\textstyle{\frac{1}{2\sqrt{2}}}(\brpsi\gamma_{\mu}\trd_{\nu}\psi
    -\trd_{\nu}\brpsi\gamma_{\mu}\psi)$ is asymmetric.
    
    \item \textbf{Gauge fields:} (Heterotic) Yang--Mills fields also couple minimally to the dilaton in DFT~\cite{Jeon:2011kp,Hohm:2011ex,Hohm:2014sxa,Choi:2015bga,Cho:2018alk} and consequently  have a non-zero dilaton charge, given by $\To=-2L_{\YM}$~\cite{Angus:2018mep}, similarly to the scalar case.  This implies that whenever the dilaton is dynamical, the fine structure constant may vary, leading to significant observational constraints \cite{Barrow:2013uza,Uzan:2010pm}.  We leave a detailed analysis of these constraints to future work, and for now content ourselves to look for solutions where the dilaton is either constant or slowly varying at late times.
    
    \item \textbf{Ramond--Ramond sector:} With an $\ODD$-symmetric unifying formulation
    ~\cite{Jeon:2012kd}, $\To=0$~\cite{Angus:2018mep}.
    
    \item \textbf{Point Particle sources:} It has been shown that $\To$ vanishes for this case. The point particle follows geodesics defined with respect to the  string-frame metric~\cite{Ko:2016dxa}. 
    
    \item \textbf{String sources:}  To zeroth order in $\alpha'$, strings do not couple to the dilaton, so their $\To$ also vanishes.  However, this should change when we include $\alpha'$ corrections~\cite{Fradkin:1984pq, Fernandez-Melgarejo:2018wpg}.

\end{itemize}

In particular, in order to study the effects of non-trivial $\To$, we will derive cosmological solutions for the cosmological constant, study the scalar case thoroughly, and consider a generalized perfect fluid. The gauge fields alone shall be considered in future works.

\subsection{Energy conditions} \label{sec:econd}
In section 4.3 of \cite{Angus:2018mep}, various energy conditions were considered in the context of  static, spherically symmetric solutions in Stringy Gravity, which are conjectured to constrain which solutions are allowed physically.  Here we re-present and extend them for the case of homogeneous and isotropic backgrounds.  Note that the expressions are simply conjectured and constructed in analogy with GR, since a full DFT description is not currently known and is beyond the scope of this work.

\begin{itemize}
	\item The \emph{strong energy condition (SEC)} is defined such that it includes magnetic $H$-flux as well as matter contributions.
	On cosmological backgrounds it reduces to the usual strong energy condition in GR plus a flux term,
	\be \label{strongcosmo}
	\rho + 3p + \frac{h^{2}e^{-2\phi}}{8\pi Ga^{6}} = (1+3w)\rho + \frac{h^{2}e^{-2\phi}}{8\pi Ga^{6}} \geq 0 \,,
\qquad
\rho+p+ \frac{h^{2}e^{-2\phi}}{16\pi Ga^{6}}
=(1+w)\rho+ \frac{h^{2}e^{-2\phi}}{16\pi Ga^{6}}\geq 0 \, .
	\ee
	Note that the flux contribution is always positive, so SEC violations in GR do not necessarily imply violations here.

	\item The \textit{positive mass condition}\footnote{In \cite{Angus:2018mep} this was labelled as the ``weak energy condition'', in heuristic analogy with GR, since it similarly pertains to only the $tt$-component, $-K_{t}{}^{t}$.   However, for consistency with our definitions \eqref{setrhop}, we hereby consider  it as another independent condition.
Furthermore,  the  inequality  $\rho+p+\frac{h^{2}e^{-2\phi}}{16\pi G a^{6}}\geq 0$ has been included in (\ref{strongcosmo}) and (\ref{WEC}) (here a strict inequality) in order to be  fully  consistent  with the standard  definitions of the energy conditions in GR.   Genuine DFT justification remains to be found, but this is beyond the scope of the present work.} depends in general on electric $H$-flux and the stringy energy-momentum tensor component $K^{t}{}_{t}$.
	Since electric $H$-flux is forbidden on cosmological backgrounds due to the requirement of homogeneity, this constraint simply becomes
	\be \label{weakcosmo}
	2\rho - \To e^{-2\phi} = (2-\lambda)\rho \geq 0 \,.
	\ee
	In \cite{Angus:2018mep} it was noted  that local violations of this condition may give rise to a regime where gravity becomes repulsive.\footnote{Technically this required a corresponding \emph{density condition}, which was defined without the spatial integral present in the positive mass condition.  However on homogeneous backgrounds the integral simply yields a constant factor, so these two conditions become identical.}

	\item The \emph{weak energy condition (WEC)}  can be defined, in consistent analogy with GR, as 
	\begin{equation}
	\rho+ \frac{h^{2}e^{-2\phi}}{32\pi Ga^{6}}\geq 0\,,\qquad
\rho+p+ \frac{h^{2}e^{-2\phi}}{16\pi Ga^{6}}
=(1+w)\rho+ \frac{h^{2}e^{-2\phi}}{16\pi Ga^{6}} > 0 \, . \label{WEC}
	\end{equation}
	For the spherical solution considered in \cite{Angus:2018mep}\footnote{See (4.78) therein.} (in the case of vanishing magnetic $H$-flux), this implies  that the Noether charge associated with time translation invariance should be non-negative. 
	
	\item The \emph{pressure condition} depends only on magnetic $H$-flux and the spatial components of the stringy energy-momentum tensor.  In a cosmological context it becomes
	\be \label{PC}
	p + \frac{h^{2}e^{-2\phi}}{32\pi Ga^{6}} \geq 0 \,.
	\ee
\end{itemize}

\begin{figure}
	\centering
	\includegraphics[width=0.8\textwidth]{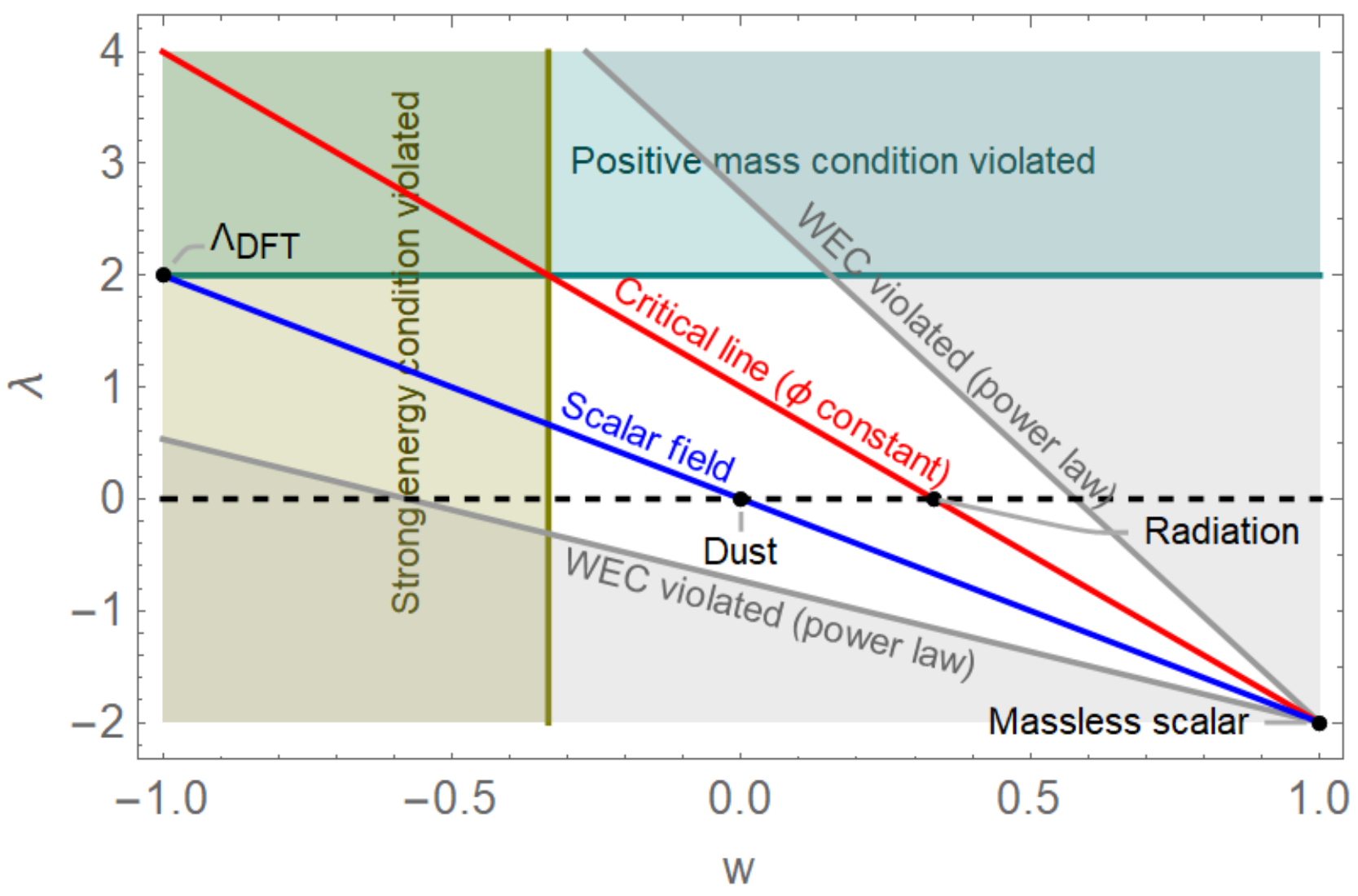}
	\caption{Energy conditions and types of matter depicted in the $(w,\lambda)$-plane.  The dotted line corresponds to conventional `SUGRA', in which the matter Lagrangian does not couple to the dilaton and hence \hbox{$\To = 0$.}  All energy conditions are respected in the white region.  The strong energy condition \eqref{strongcosmo} can in fact be preserved for $w<-1/3$ in the presence of non-trivial $H$-flux. We emphasize that the region depicted as violating the weak energy condition \eqref{WEC} is specifically restricted to the case of the power-law ansatz discussed in section~\ref{SECSolutions}.  The right-hand part of the strong~\eqref{strongcosmo} and weak energy conditions~\eqref{WEC} is automatically satisfied in the depicted region, for which $w\geq-1$.}
	\label{wlplane}
\end{figure}

\section{Solutions \label{SECSolutions}}
Having obtained the generalization of the Friedmann equations in Stringy Gravity, we turn to the important matter of finding solutions.  We first give an exposition of the general framework, before investigating examples of analytic solutions for various types of matter.

\subsection{Generalized perfect fluid}
First of all, let us consider a ``generalized perfect fluid'' in which $w$ and $\lambda$ are constant, corresponding to a linear equation of state. In cosmic gauge, the conservation equation \eqref{Conservation} is
\begin{equation}
\dot{\rho} + \left[3(1+w)H + \lambda\dot{\phi}\right]\rho = 0 \,.
\end{equation}
For constant $w$ and $\lambda$, this can be
integrated to give
\begin{equation} \label{consdensity}
\rho = \rho_{0}a^{-3(1+w)}e^{-\lambda\phi} \,,
\end{equation}
where, in keeping with usual conventions, we have defined $a_{0} \equiv 1$ and $\phi_{0} \equiv 0$.  Note that the terms in the OFEs which depend on the spatial curvature $k$ and magnetic $H$-flux $h$ can be interpreted as a particular case of $\eqref{consdensity}$: specifically, $(w,\lambda)_{h} = (1,2)$ and $(w,\lambda)_{k} = (-1/3,2)$.  Therefore in this subsection we absorb them into $\rho$ and consider a single contribution to the energy density of the form \eqref{consdensity}, which is equivalent to setting the parameters $h = k = 0$.  We will reintroduce them in the following (sub)sections when we examine specific solutions in detail.

It is instructive to consider a power law ansatz
\begin{equation} \label{powerlaw}
a = \left(\frac{t}{t_{0}}\right)^{n} \,, \qquad e^{\phi} =  \left(\frac{t}{t_{0}}\right)^{-s} \,,
\end{equation}
such that
\begin{equation} \label{powerlawHdphi}
H = \frac{n}{t} \,, \qquad \dot{\phi} = -\frac{s}{t} \, .
\end{equation}
Comparing this ansatz with the OFEs, we see that solutions with non-trivial $\rho$ are possible only if 
\be
\rho e^{2\phi}\propto t^{-2}\,,
\ee 
implying the constraint
\begin{equation}
-3n(1+ w)-s(2-\lambda)=-2 \,.\label{nsconstraint}
\end{equation}
The OFEs reduce to
\begin{align}
2\hat{\rho}_{0} &= n^{2} + 2ns + \frac{2}{3}s^{2} \, , \label{GPF1} \\
(1 + 3w)\hat{\rho}_{0} &= -n^{2} + n - ns - \frac{2}{3}s^{2} + s \, , \label{GPF2} \\
(2 - \lambda)\hat{\rho}_{0} &= -n^{2} + n + \frac{2}{3}s \, , \label{GPF3}
\end{align}
where we define a dimensionless quantity
\begin{equation} 
\hat{\rho}_{0} \equiv \frac{4\pi G}{3}\rho_{0}t_{0}^{2} \,.
\end{equation}
Note that  \eqref{GPF1}--\eqref{GPF3} are valid for any $\hrho_{0}$, while  \eqref{nsconstraint} holds only for non-vanishing $\hrho_{0}$.

Solving for $n$ and $s$ with generic $w$ and $\lambda$,  we find
\begin{equation} \label{nsgeneral}
n = \frac{2 (2w+\lambda)}{2+6w^{2}+6w\lambda+\lambda^{2}} \, , \qquad\qquad
s = \frac{2 (1-3w-\lambda)}{2+6w^{2}+6w\lambda+\lambda^{2}} \, ,
\end{equation}
and the  (not necessarily non-vanishing) energy density is proportional to
\begin{equation} \label{hatrho}
\hat{\rho}_{0} = \frac{\,6(1-w)^{2} -2(1 -3 w - \lambda)^{2}}{3(2+6w^{2}+6w\lambda+\lambda^{2})^{2}}\,.
\end{equation}
To obtain these we have made use of the fact that, from rearranging \eqref{GPF1}--\eqref{GPF3},
\begin{equation} \label{nskey}
3(1 - 3w - \lambda)\hat{\rho}_{0} = s(3n + 2s - 1) \,, \qquad 3(2w+\lambda)\hat{\rho}_{0} = n(3n + 2s - 1) \, .
\end{equation}

In order to make contact with conventional cosmology, we would like to understand the circumstances in which the dilaton $\phi$ may become constant, \textit{i.e.}~$s = 0$.  From \eqref{nsgeneral} we see that this is only possible on the \emph{critical line},
\begin{equation} \label{critline}
\lambda = 1 - 3w \, .
\end{equation}
This also holds true for more general types of matter, not just power-law solutions, as we have already seen this result in the third equation of \eqref{GRlimit}. Power-law solutions of this type exist for any $w\neq-1$, and furthermore reproduce the standard behaviour seen in cosmology based on General Relativity: plugging $s=0$ into \eqref{nsconstraint} and \eqref{hatrho} gives
\begin{equation}
n = \frac{2}{3(1+w)} \, , \qquad \hat{\rho}_{0} = \frac{2}{9(1+w)^{2}} \, .
\end{equation}
This reinforces our earlier claim that DFT cosmology encompasses all types of matter with stabilized dilaton, not just radiation as in supergravity, and thus may be the correct completion of GR.  Similarly, static solutions ($n = 0$) can only be obtained when
\begin{equation} \label{static}
\lambda = -2w \,,
\end{equation}
which is also the case for more general solutions (not just perfect fluids), and which from \eqref{nsgeneral} and \eqref{hatrho} corresponds to
\begin{equation}
s = \frac{1}{1+w} \, , \qquad \hat{\rho}_0 =
\frac{1}{3(1+w)^2} \, .
\end{equation}
Both \eqref{critline} and \eqref{static} are satisfied simultaneously when the lines intersect at $(w,\lambda)=(1,-2)$. From \eqref{nskey}, we see that in addition to the trivial solution with $n = s = 0$, this special point admits a family of solutions satisfying
\begin{equation} \label{nsfamily}
3n + 2s = 1 \, , \qquad \hat{\rho}_{0} = \frac{1}{12}(1-3n^{2}) \, .
\end{equation}
We will see in the next subsection that this corresponds to massless scalar field solutions.

We observe from \eqref{powerlawHdphi} and \eqref{nsgeneral} that for the power-law ansatz \eqref{powerlaw}, lying on the critical line\footnote{This applies to solutions on the critical line \eqref{critline}, except at $w =\pm 1$ where the denominators of \eqref{nsgeneral} and \eqref{hatrho},
\[
2 + 6w^{2} + 6w\lambda + \lambda^{2} = 3(1-w)(1+w) \, ,
\]
vanish.}
\eqref{critline} is a necessary and sufficient condition for the dilaton to be constant.  However, we emphasize that the sufficient condition applies specifically to power-law behaviour, whereas more general solutions on the critical line may have a non-trivial dilaton profile, as will be seen explicitly in some examples below.

As a final remark, note that \eqref{hatrho} is smooth as $\hat{\rho}_{0} \rightarrow 0$: in this limit it coincides with the DFT vacuum solution \cite{Copeland:1994vi} (\textit{i.e.} $\rho = 0$; see \eqref{eq:phivac} and \eqref{eq:bphivac}) with $h=k=0$.  This is a power law with
\begin{equation}
n = \pm\frac{1}{\sqrt{3}} \, , \qquad s = \frac{1}{2}(1 \mp \sqrt{3}) \, , 
\end{equation}
corresponding to the boundaries of the gray regions in Figure~\ref{wlplane}, given by $\lambda = 1\pm\sqrt{3} - (3\pm\sqrt{3})w$.

\subsection{Analytic solutions}
In order to investigate specific cases in detail, we introduce a useful gauge choice which we dub `Einstein-conformal' gauge, which is defined by the parametrization
\be
N = a \equiv be^{\phi}\,.
\ee
While the first equality fixes the time reparametrization symmetry, the second simply defines a new time-dependent function $b$, which corresponds to the Einstein-frame scale factor.  The OFEs are given in this gauge by
\begin{align}
\frac{8\pi G}{3}b^{2}e^{4\phi}\rho &= \left(\frac{b'}{b}\right)^{2} - \frac{\phi'^{2}}{3} + k - \frac{h^{2}}{12b^{4}}e^{-4\phi} \, , \label{OFE-EC1} \\
4\pi G b^{2}e^{4\phi}\rho\left(1 - w\right) &= \frac{b''}{b} + \left(\frac{b'}{b}\right)^{2} + 2k \,, \label{OFE-EC2} \\
4\pi G b^{2}e^{4\phi}\rho\left(3w + \lambda - 1\right) &= \phi'' + \frac{2b'\phi'}{b} - \frac{h^{2}}{2b^{4}}e^{-4\phi} \, , \label{OFE-EC3}
\end{align}
where $w$ and $\lambda$ were defined in \eqref{lambda}, and we have taken suitable linear combinations of \eqref{OFE1}--\eqref{OFE3} for later convenience.  In addition, the conservation equation becomes
\be \label{OFE-ECcons}
\rho' + 3\left(\frac{b'}{b}+\phi'\right)(\rho+p) + \phi'e^{-2\phi}\To = 0 \,.
\ee
For constant $w$ and $\lambda$ (\textit{i.e.} a generalized perfect fluid), this can be integrated to yield (\textit{c.f.} (\ref{consdensity}))
\begin{equation}
\rho = \rho_{0}\frac{e^{-[3(1+w)+\lambda]\phi}}{b^{3(1+w)}} = \rho_{0}\frac{e^{-\lambda\phi}}{a^{3(1+w)}} \,. \label{consdensityEC}
\end{equation}
Various types of matter will in general occupy different positions on the $(w,\lambda)$-plane, see Figure \ref{wlplane}.
However, note that in general $w$ and $\lambda$ need not be constant, for example if there are multiple competing contributions to the total energy density. 

One interesting scenario emerges on the critical line, $3w + \lambda = 1$ \eqref{critline}.  Here \eqref{OFE-EC3} can be integrated to give
\begin{equation}
(b^{2}\phi')^{2} + \frac{h^{2}}{4}e^{-4\phi} = \frac{h_{\rm{o}}^{2}}{4} \, , \label{critphih}
\end{equation}
where $h_{\rm{o}}$ is a real constant.  Plugging this back into \eqref{OFE-EC1} yields
\begin{equation}
\frac{8\pi G}{3}b^{6}e^{4\phi}\rho = \left(bb'\right)^{2} + kb^{4} - \frac{h_{\rm{o}}^{2}}{12} \, . \label{OFE-EC1crit}
\end{equation}
From \eqref{critphih} and \eqref{OFE-EC1crit}, we see that $h_{\rm{o}}^2$ represents the total energy 
shared between the dilaton and $H$-flux, which is conserved on the critical line.  In the special case where $h_{\rm{o}}$ = 0, the positive-definiteness of \eqref{critphih} forces the $H$-flux and variation of the dilaton to vanish, and hence our framework recovers the standard Friedmann equations in this limit.  More generally, we can solve for $\phi$ by casting \eqref{critphih} in the form
\begin{equation}
\frac{\rd\phi}{\sqrt{h_{\rm{o}}^{2} - h^{2}e^{-4\phi}}} = \pm\frac{\rd\eta}{2b^{2}(\eta)} \, . \label{critphiint}
\end{equation}
Here $b^{2}(\eta)$ is obtained explicitly as a function of conformal time $\eta$ by solving \eqref{OFE-EC2}, which is independent of $\phi$ on the critical line (\textit{c.f.~}\eqref{consdensityEC}).  We now discuss some explicit examples.


\subsubsection*{Pure DFT vacuum}
In Stringy Gravity, the metric $g_{\mu\nu}$ is supplemented by the additional fields $B_{\mu\nu}$ and $\phi$.  Therefore it is worth considering purely stringy  gravitational solutions, since these may be non-trivial due to possible interactions within the extended gravitational sector.  This simple scenario, which corresponds to $\rho = p = \To = 0$, can thus yield some initial insight into the nature of cosmological evolution in Stringy Gravity.  While these solutions are known in the supergravity literature (see e.g. \cite{Copeland:1994vi}), we include them here for completeness since they provide a foundation for more general solutions featuring additional matter.

In this scenario, equation \eqref{OFE-EC2} can be recast as
\begin{equation}
(b^{2})'' + 4kb^{2} = 0 \,,
\end{equation}
which has a general solution given by
\begin{equation} \label{eq:b2vac}
b^{2}=\frac{C_1\tau}{1 + k\tau^2} =\left\{\ba{lll}
\frac{C_1}{2}\sin(2(\eta-\eta_{0}))\quad&\mbox{for}&k=1\,,\\
C_1(\eta-\eta_{0})\quad&\mbox{for}&k=0\,,\\
\frac{C_1}{2}\sinh(2(\eta-\eta_{0}))\quad&\mbox{for}&k=-1\,,
\ea\right. 
\end{equation}
where $C_1$ and $\eta_{0}$ are integration constants, and we have defined
\begin{equation} \label{TauDef}
\tau =\left\{\ba{lll}
\tan(\eta-\eta_{0})\quad&\mbox{for}&k=1\,,\\
\eta-\eta_{0}\quad&\mbox{for}&k=0\,,\\
\tanh(\eta-\eta_{0})\quad&\mbox{for}&k=-1\,.
\ea\right.
\end{equation}
Consistency with \eqref{OFE-EC1} requires that the integration constants are related by
\begin{equation}
3C_1^{2} =  h_{\rm{o}}^{2} \, . \label{eq:C1h0relation}
\end{equation}
Using the fact that
\begin{equation} \label{dtaudeta}
\frac{\rmd\tau}{\rmd\eta} = 1 + k\tau^2 \, ,
\end{equation}
and using \eqref{eq:C1h0relation}, the general solution for the dilaton can be expressed as \cite{Copeland:1994vi}
\begin{equation} \label{eq:phivac}
e^{2\phi} = \left(\frac{\tau}{\tau_{\ast}}\right)^{\pm\sqrt{3}} + \frac{h^{2}}{12C_{1}^{2}}\left(\frac{\tau}{\tau_{\ast}}\right)^{\mp\sqrt{3}} \, ,
\end{equation}
where $\tau_{\ast}$ is an integration constant.
Note that this has a minimum at $\left.e^{2\phi}\right|_{\rm{min}} = |h/(\sqrt{3}C_{1})|$.  The scale factor of the original string-frame metric is thus given by \cite{Copeland:1994vi}
\begin{equation} \label{eq:bphivac}
a^2 = b^{2}e^{2\phi} = \frac{C_1\tau}{1 + k\tau^2}\left[\left(\frac{\tau}{\tau_{\ast}}\right)^{\pm\sqrt{3}} + \frac{h^{2}}{12C_{1}^{2}}\left(\frac{\tau}{\tau_{\ast}}\right)^{\mp\sqrt{3}}\right] \, .
\end{equation}

\subsubsection*{$\ODD$-symmetric DFT cosmological constant}

The action for a DFT cosmological constant in a Riemannian background is simply given by~\cite{Jeon:2011cn}
\begin{align}
S_\Lambda = -\frac{1}{8\pi G} \int d^4 x\sqrt{-g} e^{-2\phi} \Lambda \,,
\end{align}
which implies
\begin{equation} \label{eq:LambdarhopTo}
\rho_{\Lambda}e^{2\phi} = \frac{\Lambda}{8\pi G} \, , \quad\quad w_\Lambda = -1\,, \quad \quad \lambda_\Lambda = 2 \,.
\end{equation}
Note that this corresponds to a scalar field (discussed below) in the limit $\Phi' = 0$ and $V = V_{0} \equiv \Lambda/8\pi G$.

Choosing cosmic gauge ($N=1$), 
there is a solution for a static universe ($H=0$) with the dilaton evolving as
\begin{align}
\phi(t) = \pm \sqrt{\left(\frac{\Lambda}{2} - \frac{k}{a^{2}}\right)}\;t + \phi_0 \,, \qquad
k = \frac{h^{2}}{4a^{4}} \,. \label{LambdaStatic}
\end{align}
In this case, non-trivial $H$-flux implies positive spatial curvature, and physical solutions require $\Lambda a^{2} \geq 2k$ (note that $a$ must be constant).  For $h = k = 0$ and $\Lambda > 0$ this recovers the well-known static solution in flat Minkowskian spacetime  with a  linear dilaton (see e.g.~\cite{Polchinski:1998rq}):
\be
\ba{ll}
\phi(t)=\pm m(t-t_{0}) + \phi_{0}\,,\quad&\quad 
\rd s^{2}=-\rd t^{2}+\rd x^{2}+\rd y^{2}+\rd z^{2}\,,
\ea
\label{linearphi}
\ee
where $m\equiv\sqrt{\Lambda/2} > 0$.

More generally, there are expanding  solutions for $h=k=0$ and $\Lambda > 0$ with  positive $m= \sqrt{\Lambda/2}$~\cite{Mueller:1989in} given by
\be
e^{2\phi(t)}=C_{\phi}\frac{\tanh^{\sqrt{3}}\big(m(t-t_{0})\big)}{\sinh\big(2m(t-t_{0})\big)}\,,\qquad
a^{2}(t) = a_{0}^{2} \tanh^{\sqrt{\frac{4}{3}}}\big(m(t-t_{0})\big)\,,
\label{SOLL}
\ee
which is real and positive for $t>t_{0}$, and
\begin{equation}
e^{2\phi(t)} = C_{\phi}\frac{\coth^{\sqrt{3}}\big(m(t_{0}-t)\big)}{\sinh\big(2m(t_0-t)\big)}\,, \qquad
a^2(t) = a_{0}^{2}\coth^{\sqrt{\frac{4}{3}}}\big(m(t_{0}-t)\big) \, , \label{SOLL2}
\end{equation}
which is similarly defined for $t<t_{0}$ and can be obtained from \eqref{SOLL} by a combined spatial T-duality (Table~\ref{Trule}) and time-reversal transformation.
On the other hand, acting with T-duality or time reversal alone produces collapsing solutions.
Note that \eqref{SOLL} describes a decelerating expansion, whereas \eqref{SOLL2} is an accelerating solution.
The static solution \eqref{linearphi} may be derived by taking large-$t$ limits of~\eqref{SOLL} and~\eqref{SOLL2}: specifically, \eqref{SOLL} converges to the negative-sign case of~\eqref{linearphi} as $t\rightarrow\infty$, while \eqref{SOLL2} similarly converges to the positive-sign case as $t\rightarrow -\infty$, with $C_{\phi} = \half e^{2\phi_{0}}$.
For further discussion of linear dilaton solutions in DFT, see \cite{Bekaert:2016isw}.

\subsubsection*{Scalar field, \textit{e.g.} massless limit}
The action for a spatially homogeneous, canonical scalar field in a Riemannian DFT background is 
\begin{align}
S_\Phi = \int d^4 x\sqrt{-g} e^{-2\phi} \Big( -\frac{1}{2} g^{\mu\nu} \nabla_\mu \Phi \nabla_\nu \Phi - V(\Phi) \Big) \, .
\end{align}
In Einstein-conformal gauge, this yields the equation of motion
\begin{equation} \label{eq:scalarcons}
\Phi'' + \frac{2b'\Phi'}{b} + b^{2}e^{2\phi}\frac{\rmd V}{\rmd \Phi} = 0 \, .
\end{equation}
For the energy-momentum tensor components, we have
\begin{equation} \label{scalaremt}
K^t{}_t = -\frac{\Phi'^{2}}{b^2 e^{2\phi}} \, , \qquad K^r{}_r = 0 \, , \qquad \To = -\frac{\Phi'^{2}}{b^2 e^{2\phi}} + 2V(\Phi) \, ,
\end{equation}
and thus the density and pressure of $\Phi$ are given by
\begin{equation} \label{eq:rhopscalar}
\rho e^{2\phi} =  \frac{\Phi'^{2}}{2b^2 e^{2\phi}} + V(\Phi) \, , \qquad
p e^{2\phi} = \frac{\Phi'^{2}}{2b^2 e^{2\phi}} - V(\Phi) = -\frac{\To}{2}\, .
\end{equation}
We see that the equation of state is confined to the range \hbox{$-1\leq w \leq 1$} along the line
\begin{equation}
\lambda = -2w \, . \label{wlscalar}
\end{equation}

In the limit of vanishing potential, $V(\Phi) = 0$, from \eqref{eq:rhopscalar} we find that $\rho = p$ and thus $w = 1$.  In such cases \eqref{OFE-EC2} once again reduces to
\begin{equation}
(b^{2})'' + 4kb^{2} = 0 \, ,
\end{equation}
yielding again the solution \eqref{eq:b2vac}.
Furthermore, from \eqref{wlscalar} we have $\lambda = -2$, which lies on the critical line; see Figure \ref{wlplane}.  Hence we can solve for the dilaton by integrating \eqref{critphiint}, giving \cite{Copeland:1994vi,Lidsey:1999mc}
\begin{equation} \label{eq:phiscalar}
e^{2\phi} = \left(\frac{\tau}{\tau_{\ast}}\right)^{\pm\frac{h_{\rm{o}}}{C_1}} + \frac{1}{4}\frac{h^{2}}{h_{\rm{o}}^{2}}\left(\frac{\tau}{\tau_{\ast}}\right)^{\mp\frac{h_{\rm{o}}}{C_1}} \, ,
\end{equation}
where $\tau$ is as defined in \eqref{TauDef}.  This has a minimum at $\left.e^{2\phi}\right|_{\rm{min}} = |h/h_{\rm{o}}|$.
Combining this with \eqref{eq:b2vac}, the scale factor associated with the string-frame metric is given by \cite{Copeland:1994vi,Lidsey:1999mc}
\begin{equation} \label{eq:bphiscalar}
a^2 = b^{2}e^{2\phi} = \frac{C_1\tau}{1 + k\tau^2}\left[\left(\frac{\tau}{\tau_{\ast}}\right)^{\pm\frac{h_{\rm{o}}}{C_1}} + \frac{1}{4}\frac{h^{2}}{h_{\rm{o}}^{2}}\left(\frac{\tau}{\tau_{\ast}}\right)^{\mp\frac{h_{\rm{o}}}{C_1}}\right] \, .
\end{equation}

Note that for consistency with \eqref{OFE-EC1} we now require
\begin{equation}
4\pi G b^{4}\Phi'^{2} = \frac{3C_1^{2} - h_{\rm{o}}^{2}}{4} \, . \label{eq:C1h0}
\end{equation}
For real solutions, this constrains $|h_{\rm{o}}/C_1| \leq \sqrt{3}$.  The equation of motion \eqref{eq:scalarcons} gives simply $(b^{2}\Phi')' = 0$, consistent with \eqref{eq:C1h0}.  Thus the scalar field evolution is determined by the conformal Einstein scale factor $b$, with the explicit solution
\begin{equation} \label{eq:scalarsoln}
\Phi = \Phi_0 \pm \sqrt{\frac{1}{16\pi G}\left(3 - \frac{h_{\rm{o}}^{2}}{C_1^{2}}\right)}\ln\tau \, ,
\end{equation}
where we have used \eqref{eq:b2vac} and \eqref{dtaudeta}.

Finally, note that for $h = k = 0$ we have a power-law solution. 
This corresponds to the family of solutions satisfying \eqref{nsfamily}, with
\begin{equation}
n = \frac{C_1\pm h_{\rm{o}}}{3C_1\pm h_{\rm{o}}} \, , \qquad s = \frac{\mp h_{\rm{o}}}{3C_1\pm h_{\rm{o}}} \, ,
\end{equation}
as can be verified explicitly by converting to cosmic time.

\subsubsection*{Radiation solution: with $H$-flux and dynamically frozen dilaton}
For $w = 1/3$, $\lambda = 0$, we can construct a generalization of the known radiation solution in SUGRA.  In such a case 
the energy density evolves as
$\rho = \rho_{0}b^{-4}e^{-4\phi} = \rho_{0}a^{-4}$,
as expected for radiation.
From this we can write \eqref{OFE-EC2} as
\be
\left(b^{2}\right)'' + 4kb^{2} = \frac{16\pi G\rho_{0}}{3} \,,
\ee
which can be solved to give
\begin{align} \label{b2rad}
&  b^2(\eta) = \frac{\tau ( C_{1} + \cE_{0} \tau)}{1+ k \tau^2}  =  \left\{ \begin{array}{lll}
\frac{C_{1}}{2} \sin\left(2(\eta-\eta_{0})\right) + \cE_{0} \sin^2\left(\eta-\eta_{0}\right) &\mbox{for}& k=1\,, \\
C_{1}(\eta-\eta_{0}) + \cE_{0} (\eta-\eta_{0})^2   &\mbox{for}& k=0\,,\\
\frac{C_{1}}{2} \sinh\left(2(\eta-\eta_{0})\right) + \cE_{0} \sinh^2\left(\eta-\eta_{0}\right)&  \mbox{for}& k=-1 \,,\end{array} \right.
\end{align}
where $C_{1}$ is an integration constant, $\cE_{0} \equiv 8\pi G\rho_{0}/3$, and $\tau$ is as defined in \eqref{TauDef}.

Since we are on the critical line we can integrate \eqref{OFE-EC3}, 
giving \eqref{critphih}.
Applying this to \eqref{OFE-EC1} and using the solution \eqref{b2rad} for the conformal scale factor, one can verify explicitly that
\be \label{noscalars}
h_{\rm{o}}^{2} = 3C_{1}^{2} \,,
\ee
as in the vacuum solution.
Integrating \eqref{critphih} using the new scale factor \eqref{b2rad} yields\footnote{See \cite{Copeland:1994vi} for a different approach: in particular, our \eqref{eq:phirad} and \eqref{eq:bphirad} with $h = 0$ correspond to (3.26) and (3.27) therein.}
\be \label{eq:phirad}
e^{2\phi} = C_{\rm r}\left(\frac{\tau}{C_{1}+\cE_{0}\tau}\right)^{\pm\sqrt{3}} + \frac{1}{12}\frac{h^{2}}{C_{1}^{2}C_{\rm r}}\left(\frac{\tau}{C_{1}+\cE_{0}\tau}\right)^{\mp\sqrt{3}} \,,
\ee
where $C_{\rm r}$ is another integration constant.  Note that for $\cE_{0} = 0$ we should recover the vacuum solution, which implies that $\tau_{*} = C_{1}C_{\rm r}^{\mp1/\sqrt{3}}$.  Thus we can express the resulting scale factor as
\be \label{eq:bphirad}
a^{2} = b^{2}e^{2\phi} = \frac{\tau (C_{1}+\cE_{0}\tau)}{1 + k\tau^2} \left[\left(\frac{\tau}{\tau_{*}\left(1+\frac{\cE_{0}}{C_{1}}\tau\right)}\right)^{\pm\sqrt{3}} + \frac{1}{12}\frac{h^{2}}{C_{1}^{2}}\left(\frac{\tau}{\tau_{*}\left(1+\frac{\cE_{0}}{C_{1}}\tau\right)}\right)^{\mp\sqrt{3}}\right] \,.
\ee

As a bonus, we can generalize this solution to the case of radiation plus a scalar with vanishing potential (ignoring interactions).  The inhomogeneous piece of equation \eqref{OFE-EC2} depends only on the energy density in radiation, since the equivalent contribution from the scalar field vanishes, as $w_{\Phi} =1$.  Therefore $b^{2}$ must take the form \eqref{b2rad}.  Furthermore, being on the critical line, both solutions satisfy \eqref{critphih} such that the OFE \eqref{OFE-EC1} will split into a linear sum of terms for the massless scalar and radiation, respectively.  Similarly, in the non-interacting limit, the scalar field and radiation must independently satisfy the conservation equation \eqref{OFE-ECcons} (which is guaranteed by their respective equations of motion).  In all, the combined exact solution amounts to a relaxation of \eqref{noscalars}, giving 
\be \label{eq:phiradscalar}
e^{2\phi} = \left(\frac{\tau}{\tau_{*}\left(1+\frac{\cE_{0}}{C_{1}}\tau\right)}\right)^{\pm\frac{h_{\rm{o}}}{C_1}} + \frac{1}{4}\frac{h^{2}}{h_{\rm{o}}^{2}}\left(\frac{\tau}{\tau_{*}\left(1+\frac{\cE_{0}}{C_{1}}\tau\right)}\right)^{\mp\frac{h_{\rm{o}}}{C_1}} \,,
\ee
where $\tau$ is again defined by \eqref{TauDef}, and
\be \label{eq:bphiradscalar}
a^{2} = b^{2}e^{2\phi} = \frac{\tau(C_{1}+\cE_{0}\tau)}{1+k\tau^2}
\left[\left(\frac{\tau}{\tau_{*}\left(1+\frac{\cE_{0}}{C_{1}}\tau\right)}\right)^{\pm\frac{h_{\rm{o}}}{C_1}} + \frac{1}{4}\frac{h^{2}}{h_{\rm{o}}^{2}}\left(\frac{\tau}{\tau_{*}\left(1+\frac{\cE_{0}}{C_{1}}\tau\right)}\right)^{\mp\frac{h_{\rm{o}}}{C_1}}\right] \,.
\ee
Here we observe that setting $\cE_{0} = 0$ recovers the solution for a scalar field given in \eqref{eq:phiscalar} and \eqref{eq:bphiscalar}, while setting $h_{\rm{o}}^{2} = 3C_{1}^{2}$ gives the radiation solution of \eqref{eq:phirad} and \eqref{eq:bphirad}.  Taking both conditions simultaneously, we recover \eqref{eq:phivac} and \eqref{eq:bphivac}, the pure vacuum solution.
The scalar field evolution is again determined by \eqref{eq:C1h0},
however the solution to this equation now takes the form
\be \label{eq:scalarwithrad}
\Phi = \pm\sqrt{\frac{1}{16\pi G}\left(3 - \frac{h_{\rm{o}}^{2}}{C_{1}^{2}}\right)}\ln\left(\frac{\tau}{C_{1} + \cE_{0}\tau}\right) + \tilde{\Phi}_{0} \,.
\ee
Note that in the absence of radiation, $\cE_{0} = 0$, this reduces to \eqref{eq:scalarsoln} as expected. 

Consider a flat or hyperbolic universe, $k \in \{0,-1\}$, in which we may study the asymptotic behaviour at large $\eta$. In the presence of radiation, we can see from both \eqref{eq:phirad} and \eqref{eq:phiradscalar} that the dilaton tends towards a constant value and is thus dynamically frozen.  Hence  at late times this scenario recovers the standard Friedmann equations with a radiation equation of state.  Similarly, the scalar field \eqref{eq:scalarwithrad} also becomes frozen at late times.  In fact, for $k = -1$, $\tau$ itself tends towards a constant at late times (c.f.~\eqref{TauDef}), so for negative spatial curvature any purely $\tau$-dependent dilaton, such as occurs in the DFT vacuum~\eqref{eq:phivac} or massless scalar~\eqref{eq:phiscalar} cases, also freezes out.


\section{de Sitter solutions?\label{SECdeS}}
In the absence of external guidance, there lies an inevitable ambiguity in how any modified gravity sector should couple to matter --- point particles, Maxwell fields, spinor fields, any scalar fields, \textit{etc.} --- in the conventional Riemannian framework.  In a bottom-up approach, one is in principle free to choose a matter Lagrangian that is minimally coupled with respect to \textit{e.g.} the string-frame or Einstein-frame metric.  However, the stringent experimental constraints supporting the Equivalence Principle and against any ``fifth force'' require that a particle (or a planet) should follow a pure geodesic,
\be 
\label{geodesic}
\textstyle{e\frac{\rd~}{\rd\tau}}\!\left(e^{-1}\dot{x}^{\lambda}\right)+\gamma^{\lambda}_{\mu\nu}\dot{x}^{\mu}\dot{x}^{\nu}=0\,.
\ee
Furthermore, 
if the additional gravitational degrees of freedom are not stabilized, the metric appearing in the Christoffel connection, $\gamma^{\lambda}_{\mu\nu}$,
is typically expected to be that of the Einstein frame (or any sufficiently close frame).  Thus the coupling between gravity and matter should be chosen accordingly.  

From a top-down perspective, the situation is much more constrained.  The $\ODD$ symmetry principle fixes all the couplings of the closed-string massless sector to any matter.  In particular, the $\ODD$-symmetric doubled formulation of a point particle action dictates that the particle follows a geodesic with respect to the string-frame metric~\cite{Ko:2016dxa}.  On the other hand,  the $\ODD$-symmetric  action of a canonical scalar field $\chi$  reads, with the string-frame metric, 
\begin{equation}
 I_{\chi} = \int d^4 x \sqrt{-g} e^{-2\phi}\left[-\frac{1}{2}g^{\mu\nu}\partial_{\mu}\chi\partial_{\nu}\chi - V(\chi)\right]\,,
\end{equation}
which can be rewritten in terms of the Einstein-frame metric, $g^{\textrm{E}}_{\mu\nu} = g_{\mu\nu}\exp{(-2\phi)}$, as
\begin{equation}
 I_{\chi} = \int d^4x \sqrt{-g_{\rm E}} \left[-\frac{1}{2}g_{\rm E}^{\mu\nu}\partial_{\mu}\chi\partial_{\nu}\chi - e^{2\phi}V(\chi)\right]\,.
\end{equation}
Thus, in particular for a massless field, $V(\chi) = 0$, 
the dilaton completely drops out of the massless scalar action in the Einstein frame.  In accordance with \cite{Wands:1998yp}, in order to generate almost scale-invariant cosmological perturbations of $\chi$,  it might be necessary for the Einstein-frame metric $g^{\rm E}_{\mu\nu}$ to follow either an inflationary or bouncing-type evolution.  However, this is somewhat in contrast to the case of Maxwell fields,
\be
I_{{\rm{Maxwell}}} = \int d^4x \sqrt{-g} e^{-2\phi}\left[-\frac{1}{4}g^{\mu\nu}g^{\rho\sigma}
F_{\mu\rho}F_{\nu\sigma} \right]=\int d^4x \sqrt{-g_{\rm E}} e^{-2\phi}\left[-\frac{1}{4}g_{\rm E}^{\mu\nu}g_{\rm E}^{\rho\sigma}
F_{\mu\rho}F_{\nu\sigma} \right]\,,
\label{Maxwell}
\ee
where, due to the presence of a (classical) Weyl symmetry, the change of frames does not remove the dilaton.  Note that all observations are based on ordinary electromagnetic radiation and hence would be subject to the Maxwell theory of the form \eqref{Maxwell}.

In this section, we simply test whether de Sitter solutions are natural in $\ODD$-symmetric cosmology, both in string and Einstein frames separately. In order to make contact with concrete examples, here we will focus in particular on DFT coupled to a (spatially homogeneous) scalar field with arbitrary potential, which also includes the limiting case of a DFT cosmological constant.  From \eqref{scalaremt}, these solutions satisfy
\begin{equation}
K_{r}{}^{r} = pe^{2\phi}+\frac{1}{2}\To \equiv 0 \qquad \Longrightarrow \qquad \lambda \equiv -2w \, . \label{eq:dS2pT0}
\end{equation}
We will show that de Sitter solutions for such models would require exotic matter with negative energy density, violating the weak energy condition \eqref{WEC}.  Hence we may conclude that, from an $\ODD$ perspective, de Sitter is unnatural.  However, it could be interesting to investigate whether an $\ODD$ description of non-perturbative objects, such as D-branes and orientifolds \cite{Berkeley:2014nza,Berman:2014jsa,Blair:2017hhy,Blair:2018lbh,Blair:2019rrj}, can resolve this issue.

\subsection{String frame}

First of all we wish to investigate whether de Sitter solutions are allowed for the string-frame metric. 
To this end, we set $k = 0$ and consider the ansatz
\begin{equation}
a(t) = e^{Ht} \, . \label{eq:dSansatz}
\end{equation}
Here $t$ is cosmic time in string frame, which is defined by setting $N=1$.  Imposing \eqref{eq:dS2pT0} and applying \eqref{eq:dSansatz}, 
solving the OFEs \eqref{OFE1}--\eqref{OFE3} for $\dot{\phi}$ gives
\begin{equation}
\frac{\dot{\phi}}{H} = \frac{3}{2} - \frac{h^{2}}{4H^2}e^{-6Ht} \, . \label{eq:dSdphi}
\end{equation}
Inserting this into \eqref{OFE1} yields a general expression for the energy density,
\begin{equation}
\frac{8\pi G}{H^{2}}\rho e^{2\phi} = -\frac{3}{2} - \frac{h^{2}}{4H^2}e^{-6Ht} + \frac{h^{4}}{8H^{4}}e^{-12Ht} \, . \label{eq:dSrho}
\end{equation}
Further plugging \eqref{eq:dSdphi} and \eqref{eq:dSrho} into either \eqref{OFE2} or \eqref{OFE3} gives a similar expression for the pressure, 
\begin{equation}
\frac{8\pi G}{H^2}pe^{2\phi} =  -\frac{3}{2} + \frac{13h^{2}}{4H^2}e^{-6Ht} - \frac{h^{4}}{8H^{4}}e^{-12Ht} \, . \label{eq:dSp}
\end{equation}
From these we can see that as $t\rightarrow\infty$ the energy density and pressure become negative, and hence the weak energy condition~\eqref{WEC} and the pressure condition~\eqref{PC} are both violated.

Moreover, neither a DFT cosmological constant nor a scalar field is compatible with the full behaviour of~\eqref{eq:dSrho} and~\eqref{eq:dSp}.  For a non-trivial DFT cosmological constant, $w = -1$ \eqref{eq:LambdarhopTo} at all times, whereas here $w$ is varying and converges to $+1$ as $t\rightarrow\infty$. 
If on the other hand we consider a more general (canonical) scalar field, we know from \eqref{eq:rhopscalar} that when $w \rightarrow 1$ the scalar-field kinetic energy dominates.  However the energy density being negative implies that this scalar should have a wrong-sign kinetic term.

Alternatively, one might relax \eqref{eq:dS2pT0} and simply consider the special point $(w,\lambda) = (-1,4)$ on the critical line.  This indeed admits a de Sitter solution with constant dilaton, akin to the cosmological constant solution in GR. However, note that the $\ODD$-symmetric DFT cosmological constant does not correspond to $\lambda = 4$, and such a requirement does not correspond to any known $\ODD$-covariant Lagrangian.  In fact, in terms of the stringy energy-momentum tensor, such a solution would have $K_{t}{}^{t} = K_{r}{}^{r} = 4\To$, and thus in particular,
\begin{equation}
\rho = K_{t}{}^{t}e^{-2\phi} \, . \label{14solution}
\end{equation}
Since $K_{t}{}^{t}$ corresponds to minus the kinetic energy for any known type of matter, positive kinetic energy implies $\rho < 0$, so here also we expect the WEC to be violated.

\subsection{Einstein frame}
To construct the Einstein frame metric, $g_{\mu\nu}^{\textrm E} = g_{\mu\nu}\exp(-2\phi)$, it is sufficient to consider the general ansatz \eqref{CosmoAnsatz} in the gauge
\be \label{NEinstein}
N = e^{\phi} \, , \qquad a = e^{\phi}b \, .
\ee
We also define the energy density and pressure in Einstein frame as
\begin{equation}
\rho_{\textrm E} = e^{4\phi}\rho \, , \qquad p_{\textrm E} = e^{4\phi} p \, ,
\end{equation}
such that the Hamiltonian density is frame-independent: $\sqrt{-g}\rho = \sqrt{-g_{\textrm E}}\rho_{\textrm E}$.
The $\ODD$ Friedmann Equations then take the form
\begin{align}
8\pi G\rho_{\textrm E} &= 3\HE^{2} - \dot{\phi}^{2} - \frac{h^{2}}{4b^{6}}e^{-4\phi} \, , \label{OFE-E1} \\
4\pi G \left(\rho_{\textrm E} - p_{\textrm E}\right) &= \dot{\HE} + 3\HE^{2} \,, \label{OFE-E2} \\
4\pi G \left(3p_{\textrm E} - \rho_{\textrm E} + \To e^{2\phi}\right) &= \ddot{\phi} + 3\dot{\phi}\HE - \frac{h^{2}}{2b^{6}}e^{-4\phi} \, , \label{OFE-E3}
\end{align}
where we have set $k=0$ and defined the Hubble parameter in Einstein frame,
\be
\HE \equiv \frac{\dot{b}}{b} = e^{\phi}H - \dot{\phi} \, . \label{HEinstein}
\ee
Note that we have here expressed the OFEs in terms of cosmic time, $t$, in Einstein frame~\eqref{NEinstein}: if we change to conformal time, $\rd t = b\,\rd\eta$, we simply recover Einstein-conformal gauge, \eqref{OFE-EC1}--\eqref{OFE-EC3}.
The energy density and pressure in Einstein frame satisfy the conservation equation
\begin{equation}
\dot{\rho}_{\textrm E} + 3\frac{\dot{b}}{b}(\rho_{\textrm E} + p_{\textrm E}) + \dot{\phi}\left(3p_{\textrm E} - \rho_{\textrm E} + \To e^{2\phi}\right) = 0 \, .
\end{equation}

We now impose the de Sitter ansatz in Einstein frame, 
\be
b = e^{\HE t} \, . \label{EdSansatz}
\ee
Taking the difference of \eqref{OFE-E1} and \eqref{OFE-E2} yields
\be
-4\pi G \left(\rho_{\textrm E} + p_{\textrm E}\right) = \dot{\phi}^{2} + \frac{h^{2}}{4}e^{-6\HE t - 4\phi} \, , \label{EdSgeneral}
\ee
for which the right-hand side is positive-definite, implying that $\rho_{\rm E} + p_{\rm E} = \rho_{\rm{E}}(1+w) \leq 0$, violating the strong and weak energy conditions, \eqref{strongcosmo} and \eqref{WEC}, respectively.  This is the case either for $\rho_{\rm E} < 0$ and $w\geq -1$, suggesting negative-energy-density solutions as before, or $\rho_{\rm{E}}\geq 0$ and $w \leq -1$.  Here $w = -1$ is obtained only for $\dot{\phi} = h = 0$,
which in turn implies, from \eqref{OFE-E1} and \eqref{OFE-E3}, that
\begin{equation}
\rho_{\rm E} = \frac{3H_{\rm E}^{2}}{8\pi G} \, , \qquad \lambda = \frac{\To e^{2\phi}}{\rho_{\rm E}} = 4 \, . \label{GRCCE}
\end{equation}
This is again the GR cosmological-constant-like solution at $(w,\lambda) = (-1,4)$, for which the energy density is constant and dilaton-independent in Einstein frame.  However, we reiterate that there is no known $\ODD$-invariant Lagrangian corresponding to this solution, \textit{c.f.} \eqref{14solution}.


\section{Summary and discussion\label{SECCON}}
If string theory is the correct underlying description of the universe, its symmetries should constrain the types of interactions which are possible in nature.  Assuming a stringy off-shell $\ODD$ symmetry  in $D$ spacetime dimensions, the interactions between stringy gravitons (massless NS-NS sector) and matter obey the Einstein Double Field Equations \cite{Angus:2018mep}.  In this paper we studied these equations in a cosmological setting: choosing a homogeneous and isotropic ansatz for the stringy gravitons and stringy energy-momentum tensor, we obtained the $\ODD$-completion of the Friedmann equations \eqref{OFE1}--\eqref{OFE3} in the case of $D=4$.

{The $\ODD$ Principle generates a non-minimal coupling between matter and both the dilaton and the $B$-field, encoded in the stringy energy-momentum tensor as an additional scalar  component $\To$ and a skew-symmetric component $K_{[\mu\nu]}$, respectively~(\ref{EDFE}).  In particular, the presence of $\To$ justifies earlier attempts at including a dilatonic source in the SUGRA equations of motion which were based primarily on the (spatial) $\mathbf{O}(d,d)$ symmetry group present in SUGRA cosmology}.  In the perfect-fluid description, this leads to an equation of state determined by an extra parameter $\lambda$, in addition to the usual parameter $w$ denoting the ratio of pressure to energy density in a linear barotropic fluid.  The resulting two-dimensional parameter space (Figure~\ref{wlplane}) contains a critical line on which GR-like solutions with constant dilaton are admitted, which coincides with conventional SUGRA 
only at $w = 1/3$, corresponding to radiation.  This suggests that $\ODD$ cosmology may be the correct completion of both GR and SUGRA.   We also identified various solutions beyond the power-law limit, including a radiation solution with non-trivial $H$-flux and a dynamically evolving dilaton that tends towards a constant at late times.  Finally, we considered various energy conditions and found that, whether defined with respect to a string-frame or Einstein-frame metric, de Sitter solutions require negative energy density and hence violate the weak energy condition.  Thus de Sitter solutions appear unnatural in the $\ODD$ paradigm. {It may then be worthwhile to look for other types of solution to \eqref{OFE1}--\eqref{OFE3} as alternatives to de Sitter, which must nevertheless be consistent with the observed accelerating expansion of the universe.  Indeed, for some parameter ranges (\textit{e.g.~}$\tau<0$, $C_{1}<0$,  $\cE_{0}>0$), our analytic solution~(\ref{eq:bphiradscalar}) describes a bouncing universe featuring an infinite past before the bounce (hence avoiding the horizon problem) and an accelerating expansion after the bounce in cosmic time. This seems to deserve further study.}

$\ODD$ cosmology provides a new and rich framework for studying the early universe.  In this work we have only begun to scratch the surface, with many important issues remaining to be addressed.  First of all, in order to make contact with observations, it is crucial to establish how astrophysical and cosmological data should be interpreted in the $\ODD$ framework.  Point particles coupled to stringy gravity travel along geodesics defined with respect to a string-frame metric, however the subtle issue of whether string-frame or Einstein-frame descriptions are appropriate for observations remains to be resolved. Furthermore, since most data in astrophysics and cosmology are obtained from electromagnetic signals, and $\ODD$ Maxwell fields couple non-minimally to the dilaton (\textit{c.f.}~\eqref{Maxwell}), the propagation of photons on non-trivial dilaton backgrounds should be studied carefully.

In order to match observations of the Cosmic Microwave Background, the spectrum of curvature perturbations should be almost scale-invariant.  It would be interesting to investigate how this may be obtained from models of the early universe based on $\ODD$ cosmology.  Moreover, since the variation of the fine structure constant is strongly constrained by observations, models of the early universe in which the dilaton is dynamical should nevertheless yield a stabilized dilaton at late times.  Solutions such as \eqref{eq:bphirad} may be useful in realizing a dynamical mechanism of dilaton stabilization, say, in the context of flux compactification.  To this end, it may be interesting to investigate, for example, whether or not the critical line of GR-like solutions can behave as an attractor at late times.

Finally, it is crucial to investigate whether $\ODD$ is truly a symmetry of our universe at early times, and whether or not it is broken at late times.  Only further exploration will reveal the answer.\\
~\\

\section*{Acknowledgements}
The authors wish to thank all  the participants, including  Robert Brandenberger and Renato Costa,  of  the workshop, \href{https://www.doublefieldtheory.com/}{Double Field Theory: Progress and Applications, Cape Town, February 2019} for valuable comments.  SA was supported by the National Research Foundation of Korea through the grant NRF-2018R1A2B2007163. GF acknowledges financial
support from CNPq (Science Without Borders) and
JSPS Fellowship. SA and GF also wish to
thank the Yukawa Institute for Theoretical Physics and Sogang  University for their hospitality while this work was developed. The work of SM was supported by Japan Society for the Promotion of
Science (JSPS) Grants-in-Aid for Scientific Research (KAKENHI) No.
17H02890, No. 17H06359, and by World Premier International Research
Center Initiative (WPI), MEXT, Japan. He is grateful to Institut Denis
Poisson for hospitality during his stay. JHP acknowledges  the hospitality  at APCTP during the program  "100+4 General Relativity and Beyond" where part of this work was done.  KC and JHP  were  supported by  the National Research Foundation of Korea   through  the Grant 2016R1D1A1B01015196.\\
~\\

\appendix

\begin{center}
{\huge\textbf{APPENDIX}}
\end{center}
\section{Cosmological principle and stringy energy-momentum tensor in DFT \label{APPENDIXKilling}}
In this Appendix we apply the cosmological principle within the ${D=4}$  Double Field Theory framework  and derive the most general form of the stringy energy-momentum tensor which is homogeneous and isotropic.

In Double Field Theory as Stringy Gravity, we describe $D$-dimensional physics using $(D+D)$ coordinates which are gauged under an $\ODD$ symmetry (corresponding to T-duality).  Ordinary undoubled physics is recovered upon taking a $D$-dimensional section of this total space.  Up to $\ODD$ rotations we are free to write the DFT coordinates as $x^{A} = (\tilde{x}_{\nu},x^{\mu})$, where $A,B,\ldots = 1,\ldots,D+D$ are $\ODD$ indices which are raised and lowered by the $\ODD$-invariant metric
\be
\cJ_{AB}=\left(\ba{cc}0&1\\1&0\ea\right)\,.
\label{invmetric}
\ee
In this parametrization, the section condition can be expressed simply as $\tilde{\p}^{\nu} \equiv 0$, where the partial derivative is understood to act on or contract with all DFT fields.\footnote{Note that for Yang--Mills fields we must generalize this to a covariant derivative, $(\p_{A} - i\cA_{A})$, such that when we write $\cA_{A} = (\tilde{\cA}^{\nu},\cA_{\mu})$, the section condition also implies $\tilde{\cA}^{\nu} = 0$.}

The gravitational sector consists of a DFT dilaton, $d$, and a dynamical DFT metric, $\cH_{AB}$,
which can be decomposed into a pair of projectors, $P_{AB} = \frac{1}{2}(\cJ + \cH)_{AB}$ and $\brP_{AB} = \frac{1}{2}(\cJ - \cH)_{AB}$. 
Furthermore the corresponding local frame has symmetry group $\Spin(1,D-1)\times\Spin(D-1,1)$, under which the projectors can be decomposed into vielbeins $\{V_{Ap},\brV_{A\brp}\}$ as $P_{AB}=V_{Ap}V_{B}{}^{p}$ and
$\brP_{AB}=\brV_{Ap}\brV_{B}{}^{p}$, where the local Lorentz indices are raised and lowered using the metrics $\eta_{pq} = \diag(-++\dots+)$ and \hbox{$\breta_{\brp\brq} = \diag(+--\dots-)$,} respectively.

Isometries in DFT are best studied using a further-generalized Lie derivative $\fcL$, which acts on $\ODD$ vector indices as well as local $\Spin(1,D-1)\times \Spin(D-1,1)$ indices, as defined in \cite{Angus:2018mep}.
For isometries parametrized by some set of $N$ DFT vectors, $\{\zeta_{a}\}$, $a = 1,\ldots, N$, 
the further-generalized Lie derivatives of each gravitational field with respect to these DFT-Killing vectors should vanish, 
\be
\ba{lllll} 
\fcL_{\zeta_{a}}V_{Ap}=0\,,\quad&\quad
\fcL_{\zeta_{a}}\brV_{A\brp}=0\,,\quad&\quad
\fcL_{\zeta_{a}}P_{AB}=0\,,\quad&\quad
\fcL_{\zeta_{a}}\brP_{AB}=0\,,\quad&\quad
\fcL_{\zeta_{a}}d=0\,.
\ea
\label{so3VV}
\ee
This in turn implies that the DFT-Killing equations, which read
\be
\ba{ll}
P_{A}{}^{C}\brP_{B}{}^{D}(\na_{C}\zeta_{aD}-\na_{D}\zeta_{aC})= 0
\,,\quad&\quad
\na_{A}\zeta_{a}^{A}= 0\,,
\ea
\ee
should be satisfied.

In the case of spatial homogeneity and isotropy in $D=4$, we identify six  `doubled' Killing vectors,  $\xi^{M}_{a}$ (rotational, $a=1,2,3$)  and $\chi^{N}_{a}$ (translational, $a=1,2,3$), which form an algebra through the $\mathbf{C}$-bracket,
\be
\ba{lll}
\left[\xi_{a},\xi_{b}\right]_{\mathbf{C}}=\sum_{c}\epsilon_{abc}\xi_{c}\,,\quad&\quad \left[\chi_{a}, \chi_{b}\right]_{\mathbf{C}}\simeq \sum_{c} k \ \epsilon_{abc} \xi_{c}\,,\quad&\quad \left[\xi_{a}, \chi_{b}\right]_{\mathbf{C}} \simeq \sum_{c} \epsilon_{abc} \chi_{c} \,,
\ea
\ee
where, for the latter two, $\simeq$ means equal up to an exact term which is a kernel of the generalized Lie derivative. 
With  the choice of section ${\tpartial^{\mu}\equiv0}$, the   doubled Killing vectors,    $\xi^{M}_{a}=(\tilde{\xi}_{a\mu},\xi_{a}^{\nu})$,  $\chi^{N}_{a}=(\tilde{\chi}_{a\mu},\chi_{a}^{\nu})$, are given,  with the notation  $\tilde{\xi}_{a}=\tilde{\xi}_{a\mu}\rd x^{\mu}$, $\xi_{a}=\xi_{a}^{\nu}\partial_{\nu}$,    $\tilde{\chi}_{a}=\tilde{\chi}_{a\mu}\rd x^{\mu}$,   $\chi_{a}=\chi_{a}^{\nu}\partial_{\nu}$, by~\cite{Ko:2016dxa}
\begin{align}
\tilde{\xi}_{1}&=\frac{\cos\varphi}{\sin\vartheta}\left[ \frac{h r^{2}}{\sqrt{1-kr^{2}}} \rd r\right]\,, \qquad &\xi_{1}&=\sin\varphi\partial_{\vartheta}
+\cot\vartheta\cos\varphi\partial_{\varphi}\,,\nn \\    
\tilde{\xi}_{2}&=\frac{\sin\varphi}{\sin\vartheta}\left[ \frac{h r^{2}}{\sqrt{1-kr^{2}}} \rd r\right]\,, \qquad &\xi_{2}&=-\cos\varphi\partial_{\vartheta}+
\cot\vartheta\sin\varphi\partial_{\varphi}\,,
\label{KillingR}\\
\tilde{\xi}_{3}&=0\,, \qquad &\xi_{3}&=-\partial_{\varphi}\,
\nn
\end{align}
for rotations, where $h$ is constant, and
{\small{\begin{align}
\tilde{\chi}_{1} &= \frac{hr^{2}}{2} \left[ -\frac{\cos^{2}\vartheta}{\sin^{2}\vartheta}\sin\varphi \rd\vartheta + \frac{\cos^{3}\vartheta}{\sin\vartheta}\cos\varphi \rd\varphi \right]\,,  &\chi_{1} &= \sqrt{1-kr^{2}} \left[ \sin\vartheta\cos\varphi \partial_{r} + \frac{\cos\vartheta\cos\varphi}{r} \partial_{\vartheta} -\frac{ \sin\varphi}{r \sin\vartheta} \partial_{\varphi} \right] \,, \nn
\\
\tilde{\chi}_{2} &= \frac{hr^{2}}{2} \left[ \frac{\cos^{2}\vartheta}{\sin^{2}\vartheta} \cos\varphi \rd\vartheta + \frac{\cos^{3}\vartheta}{\sin\vartheta} \sin\varphi \rd\varphi \right]\,, &\chi_{2} &= \sqrt{1-kr^{2}} \left[ \sin\vartheta\sin\varphi \partial_{r} + \frac{\cos\vartheta\sin\varphi}{r} \partial_{\vartheta} +\frac{ \cos\varphi}{r \sin\vartheta} \partial_{\varphi} \right]\,, \nn \\
\tilde{\chi}_{3} &= -\frac{hr^{2}}{2} \left[ 1 + \cos^{2}\vartheta \right] \rd\varphi\,, &\chi_{3} &= \sqrt{1-kr^{2}} \left[ \cos\vartheta \partial_{r}  -\frac{\sin\vartheta}{r} \partial_{\vartheta} \right]\,
\label{KillingT}
\end{align}}}
for translations.\footnote{Note that in terms of ``Cartesian'' coordinates, $(x^{1}, x^{2}, x^{3}) = (r\sin\vartheta\cos\varphi, r\sin\vartheta\sin\varphi, \cos\vartheta)$, we have simply \\ $\chi_{i} = \sqrt{1-kx_{j}x^{j}}\p_{i}$.} 

On Riemannian backgrounds, 
the DFT-gravitational fields reduce to the closed-string (NS-NS) massless sector, $\{g_{\mu\nu}, B_{\mu\nu}, \phi\}$. Explicitly, the DFT-vielbeins separate into components as
\be
\ba{ll}
V_{Mp}=\frac{1}{\sqrt{2}}\left(\ba{c}e_{p}{}^{\mu}\\
e_{\nu}{}^{q}\eta_{qp}+B_{\nu\sigma}e_{p}{}^{\sigma}\ea\right)\,,\quad&\quad
\brV_{M\brp}=\frac{1}{\sqrt{2}}\left(\ba{c}\bre_{\brp}{}^{\mu}\\
\bre_{\nu}{}^{\brq}\breta_{\brq\brp}+B_{\nu\sigma}\bre_{\brp}{}^{\sigma}\ea\right)\,,
\ea
\label{00V}
\ee
where $e_{\mu}{}^{p}e_{\nu p}=-\bre_{\mu}{}^{\brp}\bre_{\nu\brp}=g_{\mu\nu}$, while the DFT-dilaton is expanded as
\be
e^{-2d}=\sqrt{-g}\,e^{-2\phi}\,.
\label{00d}
\ee
The above DFT-Killing equations then reduce to
\be
\ba{lll}
\cL_{\xi_{a}}g_{\mu\nu}=0\,,\quad&\qquad
\cL_{\xi_{a}}B_{\mu\nu}
+\partial_{\mu}\tilde{\xi}_{a\nu}-\partial_{\nu}\tilde{\xi}_{a\mu}=0\,,\quad&\qquad\cL_{\xi_{a}}\phi=0\,, \\
\cL_{\chi_{a}}g_{\mu\nu}=0\,,\quad&\qquad
\cL_{\chi_{a}}B_{\mu\nu}
+\partial_{\mu}\tilde{\chi}_{a\nu}-\partial_{\nu}\tilde{\chi}_{a\mu}=0\,,\quad&\qquad\cL_{\chi_{a}}\phi=0\,,
\label{00Killingeqrel}
\ea
\ee
where $\cL$ is the ordinary (Riemannian) Lie derivative.

Solving \eqref{00Killingeqrel} for the DFT-Killing vectors \eqref{KillingR} and \eqref{KillingT} in $D=4$ yields an expression for the most general (Riemannian) metric and $B$-field, 
\be
\ba{l} \label{cosmoansatz}
\rd s^{2}=-N^{2}(t)\rd t^{2} +a^{2}(t)\left[ \frac{1}{1-kr^{2}} dr^2 +r^2 d \Omega^{2}\right]\,,\\
B_{\scriptscriptstyle{(2)}}=\frac{h r^{2}}{\sqrt{1-kr^{2}}}\cos\vartheta\,\rd r\wedge\rd\varphi \,,\qquad \phi = \phi(t) \,,
\ea
\ee
where $\rd s^{2} = g_{\mu\nu}\rd x^{\mu}\rd x^{\nu}$, $\rd\Omega^{2}=\rd\vartheta^{2}+\sin^{2}\vartheta\rd\varphi^{2}$,  $B_{\scriptscriptstyle{(2)}}=\half B_{\mu\nu}\rd x^{\mu}\wedge\rd x^{\nu}$, and  
 $h$ is the same constant as in \eqref{KillingR}. 
The corresponding $H$-flux,
\be
H_{\scriptscriptstyle{(3)}}=d B_{\scriptscriptstyle{(2)}} = \frac{hr^{2}}{\sqrt{1-kr^{2}}}\sin(\vartheta) dr\wedge d\vartheta \wedge d\varphi \,,
\ee
is homogeneous and isotropic,
\be
\ba{ll}
\cL_{\xi_{a}}H_{\scriptscriptstyle{(3)}}=0 \,,\quad&\quad\cL_{\chi_{a}}H_{\scriptscriptstyle{(3)}}=0\,.
\ea
\label{HRT}
\ee

It is worthwhile to express \eqref{cosmoansatz} in terms of cartesian coordinates,
\be
\ba{l}
\rd s^{2}=-N^{2}(t)\rd t^{2} +a^{2}(t)\left[ \rd{\rm\bf x}^{2}+k\frac{\,({\rm\bf x}\cdot\rd{\rm\bf x})^{2}}{1-k{\rm\bf x}\cdot{\rm\bf x}}\right]\,,\\
B_{\scriptscriptstyle{(2)}}=\frac{h}{\sqrt{1-k{\rm\bf x}\cdot{\rm\bf x}}}\left[
z\rd x\wedge\rd y+\frac{1}{3}\left(\frac{y\rd x-x\rd y}{x^{2}+y^{2}}\right)
\wedge\rd (z^{3})\right]
=\frac{h}{\sqrt{1-k{\rm\bf x}\cdot{\rm\bf x}}}\left[
z\rd x\wedge\rd y-\frac{1}{3}\rd\tan^{-1}\!\left(\frac{y}{x}\right)
\wedge\rd (z^{3})\right]\,.
\ea
\ee
For $h=0$ and $k=0$ only, this cosmological  ansatz is preserved and the OFEs are invariant  under the entire spatial T-duality rotation given in Table~\ref{Trule},~\textit{c.f.~}\cite{Jeon:2012kd}. 
In such cases the stringy energy-momentum tensor assumes a  simple form: 
\begin{equation}
T_{AB}=\left(\ba{cc}
~0~&~\EK^{\mu}{}_{\tau}-\frac{1}{2}\To\delta^{\mu}{}_{\tau}~\\
~-\EK_{\sigma}{}^{\nu}-\frac{1}{2}\To\delta_{\sigma}{}^{\nu}~&~0~
\ea\right)\,.
\end{equation}
Note that as the metric and the corresponding vielbein  are diagonal, we have  $e_{a}{}^{\mu}=(\bre_{\mu}{}^{\bra})^{-1}$ in the diagonal gauge fixing of $\Spin(1,3)\times\Spin(3,1)$, and 
\be
\ba{ll}
K^{\mu}{}_{\mu}=e_{a}{}^{\mu}\bre_{\mu}{}^{\bra}K^{a}{}_{\bra}=K^{a}{}_{\bra}\quad&\quad ({\rm{no~\,}}\mu\,~{\rm{nor\,~}}a\,,\,\bra~\,{\rm{sum}})\,,
\ea
\ee
where $a,\bra$ are the local Lorentz vector indices which correspond to the same curved index $\mu$.

In order to couple Stringy Gravity to matter, we must introduce a non-trivial energy-momentum tensor $T_{AB}$, which appears on the right-hand side of the Einstein Double Field Equations \eqref{GenEinstein} and can be derived case-by-case from the gravitational variation of an appropriate $\ODD$-invariant matter Lagrangian \cite{Angus:2018mep}.  On homogeneous and isotropic backgrounds, we similarly require that this stringy energy-momentum tensor satisfies
\be
\fcL_{\xi_{a}}\EM_{AB}=0\,, \qquad \fcL_{\chi_{a}}\EM_{AB}=0\,. \label{KEEM}
\ee
The stringy energy-momentum tensor includes the independent components $K_{p\brq}$ and $\To$, where
\be
\ba{ll}
\EM_{AB}:=4V_{[A}{}^{p}\brV_{B]}{}^{\brq}\EK_{p\brq}-\half\cJ_{AB}\To\,.
\ea
\label{EMSG}
\ee
With \eqref{so3VV} and \eqref{EMSG}, \eqref{KEEM} decomposes into
\be
\fcL_{\xi_{a}}\EK_{p\brq}=0\,,\qquad
\fcL_{\xi_{a}}\To=0\,, \qquad
\fcL_{\chi_{a}}\EK_{p\brq}=0\,,\qquad
\fcL_{\chi_{a}}\To=0\,.
\label{Killingeqrel}
\ee
The latter two equations imply that $\To(t)$ must be at most time-dependent, while the former equations reduce to 
\be
\cL_{\xi_{a}}\EK_{\mu\nu}=0\,, \qquad \cL_{\chi_{a}}\EK_{\mu\nu}=0\,,
\ee
where we have used the convention ${\EK_{p\brq}=\half e_{p}{}^{\mu}\bre_{\brq}{}^{\nu}\EK_{\mu\nu}}$. 
This follows from the generic expression of the  further-generalized Lie derivative acting on $\EK_{p\brq}$,
\be 
\ba{ll}
\fcL_{\xi}\EK_{p\brq}=\quarter e_{p}{}^{\mu}\bre_{\brq}{}^{\nu}\!\left[2\cL_{\xi}\EK_{\mu\nu}
+\left\{2\partial_{[\mu}\tilde{\xi}_{\rho]}+\cL_{\xi}({B-g})_{\mu\rho}\right\}\!g^{\rho\sigma}\EK_{\sigma\nu}
-\left\{2\partial_{[\nu}\tilde{\xi}_{\rho]}+\cL_{\xi}({B+g})_{\nu\rho}\right\}\!
g^{\rho\sigma}\EK_{\mu\sigma}\right], \\
\fcL_{\chi}\EK_{p\brq}=\quarter e_{p}{}^{\mu}\bre_{\brq}{}^{\nu}\!\left[2\cL_{\chi}\EK_{\mu\nu}
+\left\{2\partial_{[\mu}\tilde{\chi}_{\rho]}+\cL_{\chi}({B-g})_{\mu\rho}\right\}\!g^{\rho\sigma}\EK_{\sigma\nu}
-\left\{2\partial_{[\nu}\tilde{\chi}_{\rho]}+\cL_{\chi}({B+g})_{\nu\rho}\right\}\!
g^{\rho\sigma}\EK_{\mu\sigma}\right],
\ea
\ee
together with the isometry conditions~\eqref{00Killingeqrel} and~\eqref{Killingeqrel}.

Combining these results and solving, we find that the most general form of $\EK_{\mu\nu}$ in a homogeneous and isotropic universe is diagonal, 
\be{\small{
\EK^{\mu}{}_{\nu}=\left(\ba{cccc}
~\EK^{t}{}_{t}(t)~&~0~&~0~&~0~\\
~0~&~\EK^{r}{}_{r}(t)~&~0~&~0~\\
~0~&~0~&~ \EK^{r}{}_{r}(t)~&~0~\\
~0~&~0~&~0~&~\EK^{r}{}_{r}(t)
\ea\right)\,, }}
\ee
where $\EK^{t}{}_{t}(t)$ and $\EK^{r}{}_{r}(t)$ are time-dependent functions.  Note in particular that the antisymmetric part $K_{[\mu\nu]} = 0$, which in three spatial dimensions is consistent with homogeneous and isotropic $H$-flux \eqref{HRT} under the second Einstein Double Field Equation \eqref{EDFE}.

\end{document}